# High Spatial and Spectral Resolution Observations of the Forbidden 1.707 μm Rovibronic SO Emissions on Io: Evidence for Widespread Stealth Volcanism

Imke de Pater[1,2] , Katherine de Kleer[3] , Máté Ádámkovics[4].

1. Astronomy Department, 501 Campbell Hall, University of California, Berkeley, CA 94720, USA.
2. Faculty of Aerospace Engineering, Delft University of Technology, NL-2629 HS Delft, The Netherlands.
3. Division of Geological and Planetary Sciences, California Institute of Technology, 1200 East California Boulevard, Pasadena, CA 91125, USA.
4. Lockheed Martin Advanced Technology Center, Palo Alto, CA, USA.

In memory of Donald Charles Backer

## Abstract

We present observations obtained with the 10-m Keck telescopes of the forbidden SO $a^1\Delta \rightarrow X^3\Sigma^-$ rovibronic transition at 1.707 μm on Io while in eclipse. We show its spatial distribution at a resolution of ~0.12" and a spectral resolution of R ~2500, as well as disk-integrated spectra at a high spectral resolution (R~15,000). Both the spatial distribution and the spectral shape of the SO emission band vary considerably across Io and over time. In some cases the SO emissions either in the core or the wings of the emission band can be identified with volcanoes, but the largest areas of SO emissions usually do not coincide with known volcanoes. We suggest that the emissions are caused by a large number of stealth plumes, produced through the interaction of silicate melts with superheated $SO_2$ vapor at depth. The spectra, in particular the elevated wing of the emission band near 1.69 μm, and their spatial distribution strongly suggest the presence of non-LTE processes in addition to the direct ejection of excited SO from the (stealth and other) volcanic vents.

# 1. Introduction

Io's forbidden SO 1.707 μm rovibronic transition, $a^1\Delta \rightarrow X^3\Sigma^-$, was discovered in 1999 when the satellite was observed while in eclipse (in Jupiter's shadow) with the NIRSPEC spectrometer on the Keck telescope (de Pater et al., 2002). The emission was attributed to SO molecules in the excited $a^1\Delta$ state at a rotational temperature of 1000 K, ejected from the vent at a thermodynamic quenching temperature of ~1500 K. At the time Loki Patera was suggested as its source, a volcano that was exceptionally active during that period. In subsequent years the disk-integrated SO emission was observed to vary substantially over time, in a manner not inconsistent with Loki Patera's activity (Laver et al., 2007). With a total of eight datasets, de Kleer et al. (2019a) showed that the SO total band strength across all eight dates is not correlated with incident sunlight, Io's orbital phase, the time since Io was last in sunlight, Jupiter's System III longitude, or thermal hot spot activity.

In November 2002 Io was observed with Keck's NIRSPEC spectrometer coupled to the Adaptive Optics (AO) system (de Pater et al., 2007). The authors observed Io moving through the slit to get a map of the SO distribution. They identified a latitudinal variation in SO: most emission came from the equator and the south, and practically no emission was detected in the north, despite the presence of several thermally-bright volcanic hot spots in the north.

Hence the presence of this emission band remains rather mysterious. Rotational temperatures vary from ~400 K up to ~1000 K, but in all datasets to date the wings and shoulders of the emission band could not be matched by equilibrium models at any temperature.

To further investigate the nature of the SO emission, we observed Io in eclipse with the near-infrared integral field spectrograph OSIRIS, coupled to the AO system, on the Keck II telescope in July 2010 and on Keck I in December 2015. On the latter date we observed simultaneously with the NIRSPEC spectrometer at a high spectral resolution (de Kleer et al., 2019a). We obtained one additional high spectral resolution dataset with NIRSPEC on Keck II in April 2019.

All observations are described in Section 2; the data reduction and results in Section 3, with a detailed analysis and discussion in Section 4. We end with conclusions in Section 5.

## 2. Observations

On UT 27 July 2010 and 25 December 2015 we observed Io while in eclipse with the W. M. Keck Observatory's OSIRIS (OH Suppressing InfraRed Imaging Spectrograph; Larkin et al., 2006) coupled to Keck's adaptive optics (AO) system (Wizinowich et al., 2000). OSIRIS' nominal spectral resolution is $\lambda/\Delta\lambda = 3700$ (roughly 0.5 nm). We used a platescale of 0.1"/pixel during both epochs, providing a Field-of-View (FOV) of 4.2"x6.4". The OSIRIS data reduction pipeline was used for flat-fielding, sky subtraction, cosmic ray removal, channel level adjustment and the removal of crosstalk, before extracting the spectra in the form of a data-cube (i.e., RA and DEC along the x- and y-axes; spectral channels along the z-axis). Before the data-cubes were used for science, they were rotated by the difference between the position angle of the spectrometer on the sky (PA-SPEC) and Io north on the sky[1]; in addition, the cubes were rotated by $3.6^0$ to correct for the rotation of the lenslet array relative to the dispersion axis of the grating[2]; the latter is aligned to rows on the detector. Hence, in our final image data-cubes Io north is up.

On UT 25 December 2015 and 15 April 2019 we observed Io-in-eclipse with the near-infrared spectrometer NIRSPEC (McLean et al., 1998) at its highest spectral resolution (R ~ 25,000) in the

---

[1] All ephemeris information was taken from the JPL Horizons database: https://ssd.jpl.nasa.gov/horizons.cgi#results

[2] See: https://www2.keck.hawaii.edu/inst/osiris/OSIRIS_Manual.pdf

Table 1: Summary of Observations

| Instrument | Time (UT) hr:min | Filter | Wavelength $\mu m$ | Platescale arcsec | Target | Guidestar | $\Delta^b$ AU | $r^c$ AU | Diameter arcsec | CML deg W. | Lat deg | $N \times t_{int}$ $N \times$ sec | Comments/Ref.[d] |
|---|---|---|---|---|---|---|---|---|---|---|---|---|---|
| **2010** | **27 Jul.** | | | | | | | | | | | | Eclipse ingress: $14^h:44^m$ |
| NIRC2 | 11:37-12:50 | Kcont | 2.256-2.285 | 0.01 | Io | Io | 4.367 | 4.967 | 1.155 | 315–325 | 2.35 | 2 sets of 3x15 | 1 |
| NIRC2 | 11:40-12:53 | Lp | 3.426-4.126 | 0.01 | Io | Io | 4.367 | 4.967 | 1.155 | 315–325 | 2.35 | 2 sets of 3x9 | 1 |
| NIRC2 | 11:43-12:56 | Ms | 4.549-4.790 | 0.01 | Io | Io | 4.367 | 4.967 | 1.155 | 315–326 | 2.35 | 2 sets of 3x9 | 1 |
| NIRC2 | 11:46 | Hcont | 1.569-1.592 | 0.01 | Io | Io | 4.367 | 4.967 | 316 | 1.155 | 2.35 | 3x10 | |
| OSIRIS | 13:19 | KN2 | 2.036 – 2.141 | 0.02 | Io | Io | 4.367 | 4.967 | 1.155 | 330 | 2.35 | 1x300 | |
| OSIRIS | 14:07 | KN2 | 2.036 – 2.141 | 0.035 | Io | Io | 4.367 | 4.967 | 1.155 | 336 | 2.35 | 1x300 | $SO_2$-ice |
| OSIRIS | 14:21 | HN4 | 1.652 – 1.737 | 0.10 | Io | Io | 4.367 | 4.967 | 1.155 | 338 | 2.35 | 1x2 | Io in sunlight |
| OSIRIS | 14:47-15:31 | HN4 | 1.652 – 1.737 | 0.10 | Io | Callisto | 4.367 | 4.967 | 1.155 | 342-348 | 2.35 | 32x15 | Io in eclipse Separation[a]35–29" |
| **2010** | **28 Jul.** | | | | | | | | | | | | Io in sunlight |
| NIRC2 | 11:55 | Kcont | 2.256-2.285 | 0.01 | Io | Io | 4.349 | 4.961 | 1.160 | 161 | 2.36 | 3x15 | 1 |
| NIRC2 | 11:58 | Lp | 3.426-4.126 | 0.01 | Io | Io | 4.349 | 4.961 | 1.160 | 161 | 2.36 | 3x4.5 | 1 |
| NIRC2 | 12:01 | Ms | 4.549-4.790 | 0.01 | Io | Io | 4.349 | 4.961 | 1.160 | 161 | 2.35 | 2 sets of 3x9 | 1 |
| NIRC2 | 12:06 | Lp | 3.426-4.126 | 0.01 | Io | Io | 4.349 | 4.961 | 1.160 | 162 | 2.36 | 3x9 | 1 |
| OSIRIS | 12:27-12:39 | KN2 | 2.036 – 2.141 | 0.035 | Io | Io | 4.349 | 4.961 | 1.160 | 165-167 | 2.36 | 1x300 | $SO_2$-ice |
| **2015** | **25 Dec.** | | | | | | | | | | | | Eclipse ingress: $13^h:45^m$ |
| NIRC2 | 12:06-12:26 | Kcont | 2.256-2.285 | 0.01 | Io | Io | 5.152 | 5.417 | 0.979 | 326–329 | -1.83 | 2 sets of 3x15 | 2,3[e] |
| NIRC2 | 12:08-12:42 | Lp | 3.426-4.126 | 0.01 | Io | Io | 5.152 | 5.417 | 0.979 | 326–331 | -1.83 | 2 sets of 3x18 | 2,3[e] |
| NIRC2 | 12:12-12:45 | Ms | 4.549-4.790 | 0.01 | Io | Io | 5.152 | 5.417 | 0.979 | 327–332 | -1.83 | 2 sets of 3x18 | 2,3[e] |
| NIRC2 | 12:30 | Hcont | 1.569-1.592 | 0.01 | Io | Io | 5.152 | 5.417 | 0.979 | 329 | -1.83 | 3x12 | |
| OSIRIS | 13:10-13:30 | HN4 | 1.652 – 1.737 | 0.10 | Ganymede | Io | 5.152 | 5.417 | 1.407 | 355 | -1.68 | 4x30 | Ganymede in eclipse |
| OSIRIS | 13:42 | HN4 | 1.652 – 1.737 | 0.10 | Io | Io | 5.152 | 5.417 | 0.979 | 340 | -1.83 | 1x30 | Io in sunlight |
| OSIRIS | 13:48-14:58 | HN4 | 1.652 – 1.737 | 0.10 | Io | Ganymede | 5.152 | 5.417 | 0.979 | 340-350 | -1.83 | 44x30 | Io in eclipse Separation[a]7.3–5" |
| NIRSPEC | 13:49-14:47 | NIRSPEC6 | 1.694 - 1.717 | 0.72×0.194 | Io | – | 5.152 | 5.417 | 0.979 | 340–349 | -1.83 | ~ 20x120 | 12" slit; Io in eclipse; ref. 4 |
| **2019** | **15 Apr.** | | | | | | | | | | | | Eclipse ingress: $13^h:54^m$ |
| NIRC2 | 10:36-10:38 | Lp | 3.426-4.126 | 0.01 | Io | Io | 4.735 | 5.320 | 1.056 | 314 | -2.84 | 3x20 | 2,3[e] |
| NIRC2 | 10:39-10:42 | Ms | 4.549-4.790 | 0.01 | Io | Io | 4.735 | 5.320 | 1.056 | 314 | -2.84 | 3x20 | 2,3[e] |
| NIRC2 | 10:46-10:48 | Kcont | 2.256-2.285 | 0.01 | Io | Io | 4.735 | 5.320 | 1.056 | 315 | -2.84 | 3x20 | 2,3[e] |
| NIRC2 | 10:59 | Hcont | 1.569-1.592 | 0.01 | Io | Io | 4.735 | 5.320 | 1.056 | 317 | -2.84 | 3x20 | 2,3[e] |
| NIRSPEC | 13:58-14:58 | NIRSPEC5 | 1.680 - 1.731 | 0.72×0.129 | Io | – | 4.735 | 5.320 | 1.056 | 342-350 | -2.84 | 20x120 | 12" slit; Io in eclipse |

[a] Separation between Io and guidestar at the beginning and end of eclipse indicated
[b] Geocentric distance
[c] Heliocentric distance
[d] References: 1: de Pater et al., 2014; 2: de Kleer et al., 2016; 3: de Pater et al., 2017; 4: de Kleer et al., 2019a.
[e] Narrow-band filter images were also taken in the PAH, $H_2O$-ice, $Br_\alpha$, and $Br_{\alpha-cont}$.

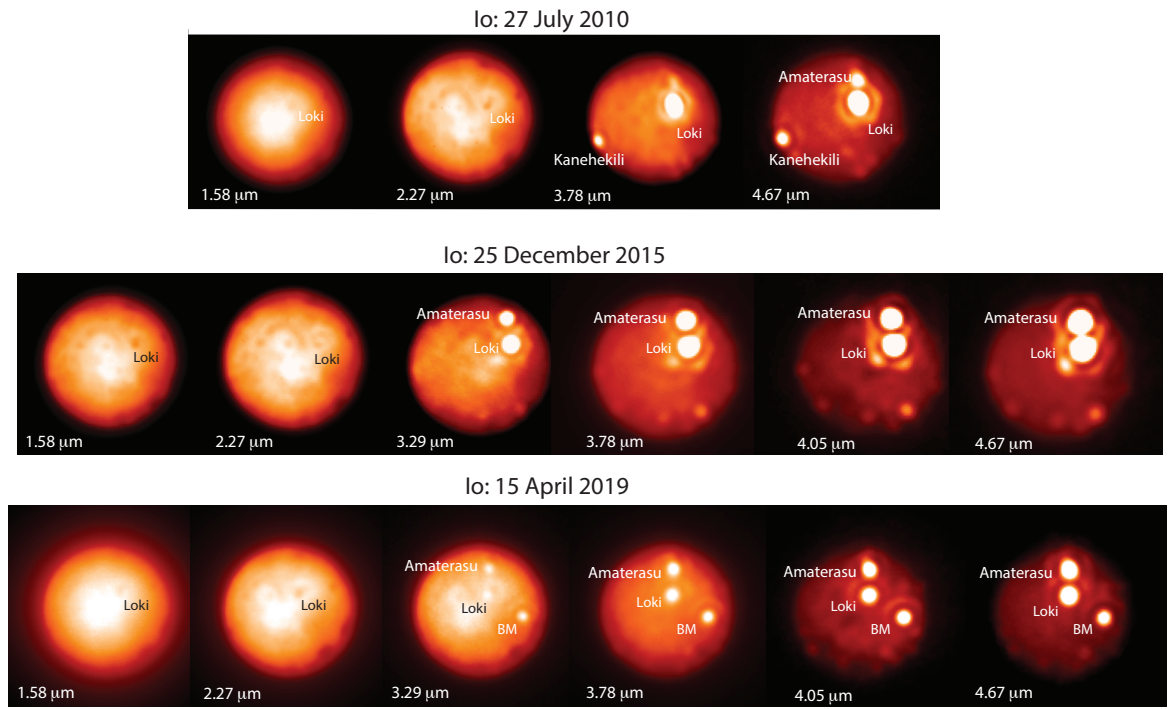

Fig. 1

Fig. 1. AO-corrected images of Io at different wavelengths (indicated on each frame) taken with NIRC2 before Io went into eclipse. The brightness contrast in each image is optimized to show both faint and bright sources. The rings around bright sources are artifacts, produced by the PSF of the telescope (i.e., Airy ring). Prominent volcanoes are indicated by name. In 2019, BM stands for Boösaule Montes, to the NW of Pele. Although the Hcont (1.58 μm) images on 27 July 2010 and 15 April 2019 are of rather poor quality, they could still be used for photometric calibration. Some of the images in this figure were also shown in de Kleer & de Pater (2016), de Kleer et al. (2019a), de Pater et al. (2014 & 2017). In all images in this figure, as well as in all other figures in this paper, Io North is up.

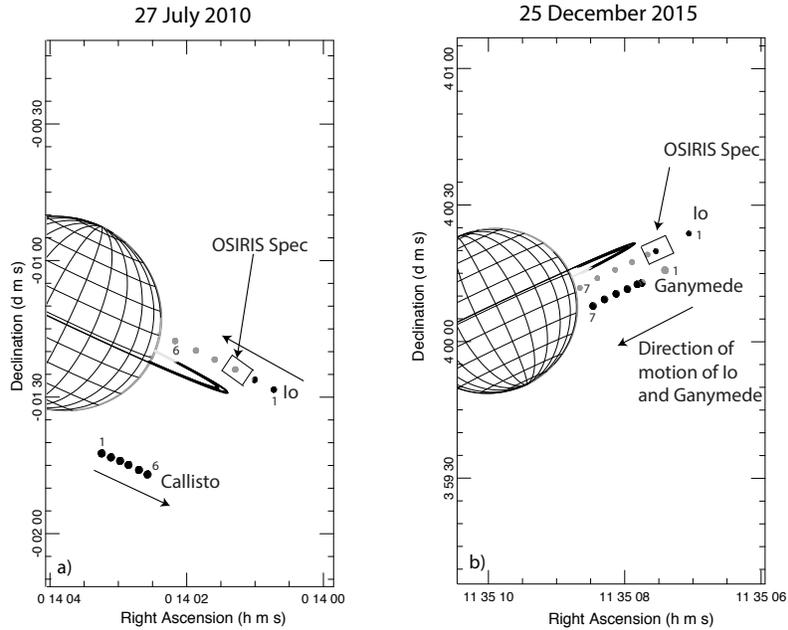

Fig. 2

Fig. 2. a) Geometry of our set-up for the OSIRIS observations in July 2010. The positions of Io and Callisto are shown at 6 different times, separated by 15 min; the first (at 14:30 UT) and last (at 15:45 UT) are indicated by 1 and 6. While Io is moving towards Jupiter, and getting into eclipse at 14:45 (grey vs black shows in-eclipse vs in-sunlight, respectively), Callisto is moving away from Jupiter, as indicated by the arrows. The approximate size and orientation of the spectrometer's field-of-view is indicated by the rectangle. b) Geometry of our set-up for the OSIRIS observations December 2015. The positions of Io and Ganymede are shown at 7 different times, the first (at 13:30 UT) and last (at 14:55 UT) are indicated by 1 and 7. Both satellites move in the same direction, indicated by an arrow. Initially, Ganymede was in eclipse; at 13:44 (step 2) Ganymede is coming out of eclipse; at 13:55 (time step 3) Io has entered Jupiter's shadow; subsequent time steps are at 14:10, 14:25, 14:40, and 14:55. (Both panels are adapted from the Planetary Ring Node; http://pds-rings.seti.org/tools/).

0.72" slit. These observations were performed under normal seeing conditions (no AO).

During each night we also obtained several images of Io with the near-infrared camera NIRC2 coupled to Keck's AO system. NIRC2 is a 1024×1024 Aladdin-3 InSb array, which we used in its highest angular resolution mode, i.e., the NARROW camera at 9.94 ± 0.03 mas per pixel (de Pater et al., 2006). These images were taken when Io was in sunlight, typically ~1-3 hrs before eclipse ingress.

A summary of all observations, including details on the timing with respect to eclipse ingress, is provided in Table 1; observing specifics are provided in each subsection below.

### 2.1 UT 27 July 2010

On UT 27 July 2010, the field-integral spectrometer OSIRIS and the near-infrared camera NIRC2, both coupled to the AO system, were on the Keck II telescope. Before Io went into eclipse, we took two sets of images of the satellite with NIRC2. Photometric calibration was performed using the nearby standard star

HD1160[3]. The images taken with the Kcont (2.27 μm), Lp (3.78 μm), and Ms (4.68 μm) filters were published by de Pater et al. (2014); the Hcont filter (1.58 μm, extending over 1.5688 – 1.5920 μm) image is used in this paper for calibration purposes (Section 3). Several images are shown in Figure 1.

At 13:00 UT we switched instruments from NIRC2 to OSIRIS. While Io was still in sunlight, we took spectra in the KN2 band (2.036 – 2.141 μm) to image the $3\nu_1+\nu_3$ $SO_2$-ice band on Io (Schmitt et al., 1994). A second $SO_2$-ice band image data-cube was obtained one day later, on 28 July 2010. The pixel scale (and FOV) for these KN2 data varied between 0.02" (FOV: 0.90"x1.28") and 0.035" (FOV: 1.58x2.24"), at a spectral resolution, R ~ 2,500. One exposure was taken on Io, and one on the sky nearby (i.e., completely off Io), each 300 sec long.

Before Io went into eclipse, we took an image data-cube of Io-in-sunlight with the HN4 filter (1.652 – 1.737 μm) at R ~ 2,500 and a pixel size of 0.1"; we used Io itself for wavefront sensing. We used this same setup when Io was in eclipse, except that we used Callisto for wavefront sensing. Callisto was 35" away from Io at the start of the eclipse, and moved 6" towards Io during the observing sequence (Fig. 2a). The long axis of the OSIRIS FOV was ~10° inclined relative to Io's direction of motion. We manually performed the variable off-set guiding: right before starting the exposure we calculated what the offset would be about a minute into the future (each 15-sec exposure took 1-1.5 min in real time), so that we could offset the telescope by the correct amount to get Io in the FOV.

We obtained 32 image data-cubes of Io-in-eclipse, each with an integration time of 15 seconds. Observations started when Io went into eclipse, at 14:48 UT, and continued until 15:31 UT, ~25 min before the end of the eclipse, at which time we switched back to NIRC2 to take a few images of Io-in-eclipse. Unfortunately, no reliable NIRC2 images were obtained during the eclipse period. OSIRIS sky frames were taken at the beginning, near the middle and end of the observing run. During the 43 min we observed Io, the satellite rotated 6 deg in longitude, which is ~0.06" at the center of Io's disk, i.e., about half a pixel size. Due to the differential motion between Io and Callisto, Io moved (i.e., was smeared) over 0.12" on the sky during the 15-sec exposure, which is roughly 1 pixel in our observations. These effects were ignored.

*2.2 UT 25 December 2015*

On UT 25 December 2015, OSIRIS was on the Keck I telescope, while the NIRC2 camera and NIRSPEC spectrometer were on the Keck II telescope. Both telescopes were used to observe Io before and while in eclipse. On Keck II we imaged Io with NIRC2 in 8 different filters between 1.6 and 5 μm (Fig. 1; see also de Kleer and de Pater, 2016; de Pater et al., 2017; de Kleer et al., 2019b). Like for the 2010 data, the Hcont image was used for flux calibration purposes. Photometric calibration was performed on the nearby standard star HD22686[3].

When Io was in eclipse, the satellite was observed with OSIRIS on Keck I at a medium spectral resolution (R ~ 2,500), and with NIRSPEC on Keck II in high spectral resolution mode (R ~ 15,000). The NIRSPEC data are presented in de Kleer et al. (2019a). As in July 2010, the SO OSIRIS spectra were taken with the HN4 filter. One image data-cube was taken before Io went into eclipse. After the satellite entered Jupiter's shadow Ganymede was used for wavefront sensing. Coincidentally, Ganymede came out of eclipse almost the moment Io went into eclipse. To be precise, Ganymede entered partial eclipse during egress at $13^h:38^m$, and was completely illuminated by $13^h:47^m$. Io entered partial eclipse during ingress at $13^h:45^m$, and by $13^h:48^m$ Io was completely in Jupiter's shadow. Before Ganymede came out of eclipse, we took a few exposures of Ganymede-in-eclipse, while using

---

[3] https://www2.keck.hawaii.edu/inst/nirc/Elias_standards.html

Io for wavefront sensing. These data were also taken in the HN4 filter, and are summarized in Section 3.4. During the entire Io-in-eclipse observing period, the two satellites remained very close together (5—7.3"). Figure 2b shows the viewing geometry and position of the FOV for the 2015 date; to optimize efficiency, the long axis of the FOV was oriented along Io's direction of motion, so the satellite drifted through the FOV. As in 2010, we manually performed the variable off-set guiding. Sky frames were taken at the beginning and end of the observing run.

We obtained 44 image data-cubes of Io-in-eclipse, each with an integration time of 30 seconds. Observations started when Io went into eclipse, at 13:48 UT, and continued until 14:57, right when Io disappeared behind Jupiter. Io rotated almost 10 degrees during this period, which induces a rotational smearing at the center of the satellite of 0.09", or just about 1 pixel. The differential motion between Io and Ganymede resulted in a smearing of only 0.05" during each 30-sec scan. These effects were ignored.

*2.3 UT 15 April 2019*

On UT 15 April 2019 we observed Io-in-eclipse with the NIRSPEC spectrometer in its high spectral resolution mode on the Keck II telescope, soon after the spectrometer (an echelle spectrograph) had undergone a major upgrade. This upgrade allowed for observations over a somewhat broader bandpass per order than previously possible. As discussed in de Kleer et al. (2019a), in December 2015 spectra were taken over the wavelength range ~1.694 – 1.717 μm, while in 2019 we covered the range 1.680 to 1.731 μm. As in December 2015, we used a slit with a length of 12" and pixel size of 0.129"; we choose the broadest possible slit-width, which is 0.72". This results in a spectral resolution of R~15,000 (McLean et al., 1998). The total integration time per Io-in-eclipse spectrum was 120 sec, and the airmass was close to 1.4 both for Io and the calibrator observations. Analogous to our OSIRIS observations, we manually performed variable off-set guiding to keep eclipsed-Io on NIRSPEC's slit. Sky frames were taken at the beginning and end of the observing run.

Before Io went into eclipse, we took spectra of the satellite while in sunlight to test our procedure of manually updating Io's tracking rate to keep the slit on the satellite while integrating; these spectra were later used for calibration purposes (Section 3.3). We continuously took images of the satellite with the slit-viewing camera, SCAM, to verify that Io was still on the slit (both in sunlight and in eclipse).

At the beginning of our night, we observed Io with NIRC2 in 9 different filters between 1.6 and 5 μm (Fig. 1). The Hcont image was used for flux calibration purposes, as for the OSIRIS data. Photometric calibration was performed using the nearby standard star BS 6441[4].

## 3. Data Reduction & Results

*3.1 NIRC2 Observations*

Three images were obtained with NIRC2 in each filter, one in each of 3 quadrants on the detector (the lower left quadrant, which has many artifacts/bad pixels, was not used). Median averaging provides a sky background.

All images were processed using standard near-infrared data reduction techniques (flat-fielded, sky-subtracted, with bad pixels replaced by the median of surrounding pixels). The geometric distortion in the Keck images was corrected using the "dewarp" routines provided by Brian Cameron of the California Institute of Technology for the 2010 and 2015 data[5], and the solution provided by Service et al. (2016) for the 2019 data. The individual images were aligned and co-added to increase signal-to-noise ratio (SNR). Images of Io at different near-infrared wavelengths are shown in Figure 1. While the

---

[4] https://www.gemini.edu/sciops/instruments/nearir-resources/photometric-standards/ukirt-bright-standards

[5] http://www2.keck.hawaii.edu/inst/nirc2/forReDoc/post-observing/dewarp/nirc2dewarp.pro

Hcont images have been used to calibrate the OSIRIS and NIRSPEC data as discussed below, the other images show which volcanoes are active, useful for comparison with the spatial distribution of the SO emissions. The precise location of these volcanoes, as measured from the NIRC2 images, was used to determine both the center of Io's disk on OSIRIS image data-cubes, and Io's orientation on the sky when the satellite was in eclipse and its limb could not be seen.

*3.2 OSIRIS SO Observations*

After the individual frames of each data-cube (32 in 2010, 44 in 2015) were inspected and bad pixels removed by replacing them with the median of the 8 surrounding pixels, the image data-cubes needed to be aligned since each data-cube was located at a slightly different position on the FOV (Sections 2.1, 2.2). We used the brightest thermal source, Loki Patera, visible in each individual data-cube, to align all of them. The data-cubes were then averaged on each day to provide a final spectral image data-cube of Io-in-eclipse. Since Io is so close to Jupiter, and in 2015 also very close to Ganymede, we modeled the background by fitting polynomials (2-degree) to each row on each image plane (e.g., each wavelength) of the data-cube, after "blocking out" Io. In 2015, we also needed to fit polynomials (1-degree) to each column, presumably due to the proximity of Ganymede. These background image data-cubes were subtracted from the data. The spatial resolution in our final image data-cubes, as determined from scans through Loki Patera, was ~0.12".

The image data-cubes were calibrated with photometrically calibrated Hcont images of Io while in sunlight, obtained with NIRC2 on the same nights (Sections 2, 3.1). At wavelengths corresponding to the Hcont filter, Io's total intensity can be attributed entirely to reflected sunlight (Fig. 1). The total flux density from Io in the Hcont image in July 2010 was 6.4 x $10^{-8}$ ergs s$^{-1}$ cm$^{-2}$ μm$^{-1}$; in December 2015 it was 3.9 x $10^{-8}$ ergs s$^{-1}$ cm$^{-2}$ μm$^{-1}$. The difference in intensities is caused by the fact that Io was much further from both the Sun and the Earth in Dec. 2015 (Table 1); the ratio $[r\Delta(2015)/r\Delta(2010)]^2 = 1.655$, which is equal to the ratio in the Hcont flux density between 2015 and 2010. These measured intensities scale with the total counts in an OSIRIS image of Io in the HN4 band while the satellite was still in sunlight. The latter image was constructed by averaging an image data-cube of Io-in-sunlight over wavelength. Spencer et al.'s (2004) spectrum of Io shows that the reflectivity of the satellite does not vary much with wavelength between 1.57 (shortest wavelength in the Hcont filter) and 1.74 μm (longest wavelength in the HN4 filter). Assuming no variation in reflectivity, we multiplied the total flux in the Hcont image by 0.80 to account for the decrease in solar flux from 1.58 to 1.695 μm, the central wavelengths in each band. We then used these values to calibrate the OSIRIS HN4 spectra. The spectra were further converted into photons s$^{-1}$ cm$^{-2}$ μm$^{-1}$. In order to compare the total flux densities in our two datasets with each other and with previous observations of SO, we scaled the SO disk-integrated spectral data-cubes (with units in photons s$^{-1}$ cm$^{-2}$ μm$^{-1}$) to correspond to a geocentric distance of 4.08 AU, the distance of Io at the time of the very first detection, 24 September 1999 (de Pater et al., 2002). Since the SO emission does not depend on solar insolation, we only scaled the intensities with the difference in geocentric distance (squared). The units on the disk-resolved data-cubes are presented in units of photons s$^{-1}$ cm$^{-2}$ sr$^{-1}$ μm$^{-1}$.

Since the SO emission band is in a wavelength region where telluric emissions/absorptions are essentially absent, as shown from observations of A stars for telluric correction, the data did not need to be corrected for telluric emissions/absorptions, which helped keep the SNR as high as possible.

A fully calibrated disk-integrated spectrum for both epochs is shown in Fig. 3, top-row, with superposed blackbody curves for a temperature of 550, 675, and 800 K. In both years, the background is matched quite well by a 675-K blackbody; in 2010 the area of this high temperature is ~50 km$^2$; in 2015 it is ~120 km$^2$. The middle row shows the disk-integrated spectrum after subtraction of these blackbody curves. These spectra, as all (OSIRIS and

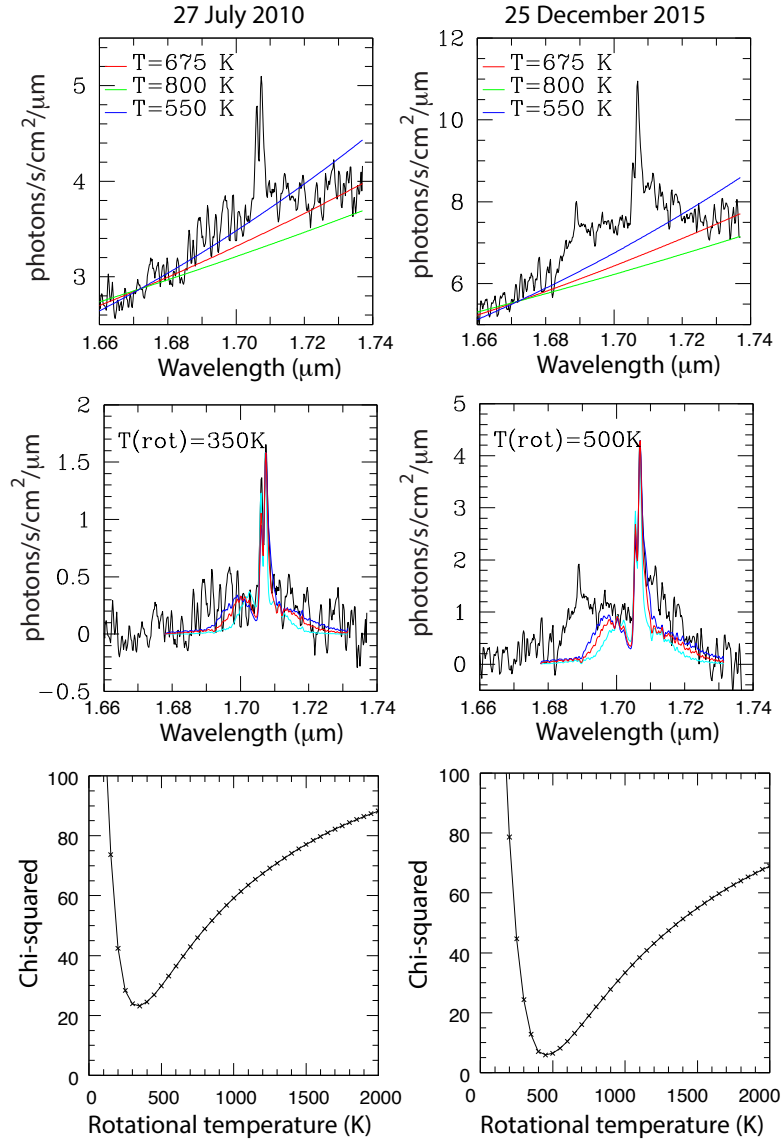

Fig. 3

Fig. 3. OSIRIS spectra of Io, integrated over the entire disk. Top row shows the spectra with superposed blackbody curves with temperatures of 550, 675, and 800 K. Middle row shows the emission after subtraction of a 675-K blackbody curve. Note that all disk-integrated spectra were scaled to a distance of 4.08 AU, so that differences in intensity between 2010 and 2015 are true variations, and not caused by changes in distance. Superposed on the middle row are LTE model spectra (discussed in Section 4) for a rotational temperature as indicated (red line), as well as lines for ± 200 K temperatures (blue: +200 K; cyan: -200 K). The best fit spectra were determined using chi-square fits to the center portion of the line; the chi-square curves as a function of rotational temperature are indicated in the bottom row. Units are in photons/s/cm$^2$/μm; one can convert these to steradians by dividing by Io's solid angle (at 4.08 AU).

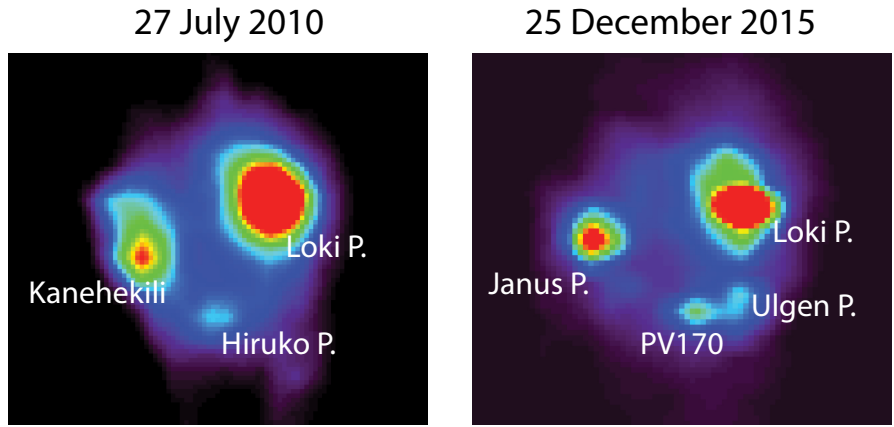

Fig. 4

Fig. 4. Images of Io-in-eclipse integrated over the entire wavelength band (HN4 filter), which is dominated by hot spot thermal emission. The colors are chosen such as to enhance the contrast.

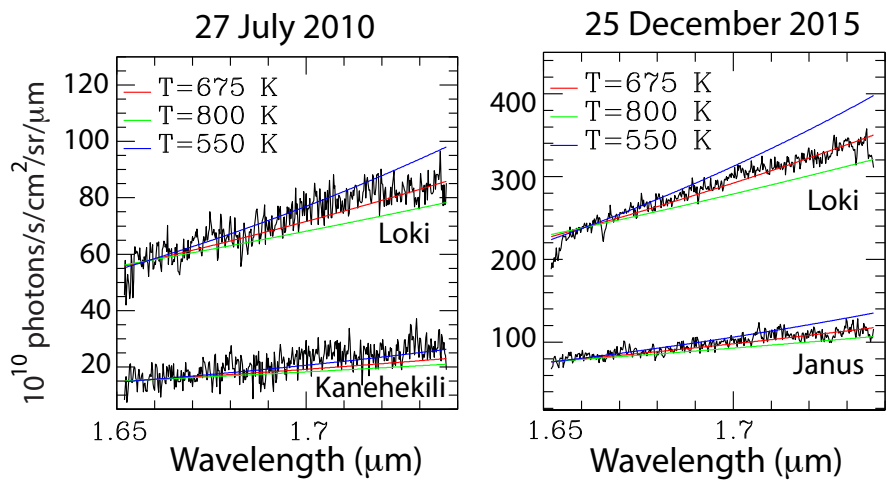

Fig. 5

Fig. 5. One-pixel spectra of Loki Patera and Kanehekili Fluctus on 27 July 2010, and of Loki Patera and Janus Patera on 25 December 2015. Superposed are blackbody curves for temperatures of 550, 675 and 800 K.

NIRSPEC) spectra in this paper, are Hanning smoothed over 5 pixels[6]. Note that both the background flux (top row) and line strength (middle row) was a factor of 2-3 higher in 2015 than in 2010, after the data were scaled to an Earth-Io distance of 4.08 AU. Hence these differences are intrinsic to Io, or in other words the SO emission was much stronger in 2015 than in 2010.

Figure 4 shows images of Io-in-eclipse, averaged over the entire HN4 filter; clearly, the total emission is dominated by thermal emission from Io's volcanoes. The total intensity of Io-in-eclipse in 2010 is $4.6 \times 10^{-12}$ ergs s$^{-1}$ cm$^{-2}$ μm$^{-1}$, which is ~0.01% of Io's intensity in sunlight. In 2015 we measure an intensity of $8.1 \times 10^{-12}$ ergs s$^{-1}$ cm$^{-2}$ μm$^{-1}$, or 0.026% of sunlit Io. As shown, in both years Loki Patera was the brightest hot spot. Judging from the periodic brightenings observed at Loki Patera (de Kleer & de Pater, 2016; de Pater et al., 2017), in both years the volcano was at the beginning of a brightening phase. Although the low spatial resolution of the images precludes an accurate determination of Loki Patera's intensity at 1.7 μm, a rough estimate gives ~0.4 GW/sr/μm in 2010, and ~1 GW/sr/μm in 2015 (this may be contaminated with flux from Amaterasu Patera; see de Kleer & de Pater, 2016). These numbers should be increased by ~20-25% due to the emission angle (foreshortening) effect, resulting in ~0.5 GW/sr/μm in 2010, and ~1.2 GW/sr/μm in 2015. For comparison, in 2015 an intensity of $1.1 \pm 0.2$ GW/sr/μm was measured at 2.2 μm (de Pater et al., 2017).

In order to assess the spatial distribution of SO gas over Io, we have to subtract the satellite's continuum emission as accurately as possible. We therefore created background images from the spectra at both epochs that do not contain SO emissions, i.e., we averaged over wavelengths shorter than 1.68 μm and longer than 1.73 μm. These images, which look similar to those shown in Fig. 4, need to be scaled with wavelength to remove the continuum emission in the HN4 filter. We used a 1-pixel spectrum of Loki Patera at its peak intensity (Fig. 5; note that the SO emission band is essentially absent in a 1-pixel spectrum since it is hidden in the noise of Loki Patera's thermal flux density) to determine a blackbody curve that matches this spectrum; these curves are superposed on the spectra in Figure 5 for both 2010 and 2015. A good match to the data is provided by a blackbody curve of 675 K in both years, just like for the disk-integrated spectra (Fig. 3). This is not surprising, since Io's thermal flux density is dominated by Loki Patera. The same blackbody spectrum also matches the slope of a 1-pixel spectrum of Janus Patera in December 2015, as shown in Figure 5. After Loki Patera, the brightest source in July 2010 was Kanehekili Fluctus. Although a lower (~550 K) temperature provides a better fit to this much fainter volcano, scaling the background with a slope in wavelength given by the 675 K profile will eliminate most of Kanehekili's thermal emission. We note that the brightness temperature results for Janus Patera and Kanehekili Fluctus agree well with previous measurements (de Pater et al., 2014b). Other volcanoes are too faint to determine a blackbody temperature.

We thus scaled the background image data-cube for each year with the slope of these Loki Patera spectra, and subtracted the scaled background image data-cube from the spectral data-cubes. Figure 6 shows the results after combining (integrating) the image data planes within the narrow core of the SO emission band (panels b, e: 1.705-1.709 μm), and within the wings of the emission band (panels c, f: 1.686-1.720 μm minus the core at 1.705-1.709 μm). Panels a and d show the emission integrated over the entire emission band (1.686-1.720 μm) before removing the background emission; as shown, the latter images look very similar to those in Figure 4. Clearly, the SO emission is only a very small fraction of the total emission, and shows no obvious correlation with the bright hot spots. We further notice that, although the emissions in the core of the line broadly agree with those in the wings, that there are some distinct differences as well. This will be discussed in more detail in Section 4.

---

[6] Spectra were Hanning smoothed as follows:
[F(i)+0.5*(F(i-1)+F(i+1))+0.25*(F(i-2)+F(i+2))]/2.5, with F the intensity at pixel i.

We superposed a circle outlining the satellite itself on each of the panels in Figure 6. Since we cannot see the limb of Io-in-eclipse, we used the volcanoes in Figure 4 with the position as determined from NIRC2 images during these epochs (Fig. 1; de Pater et al., 2014 and de Kleer et al., 2019b). We estimate the uncertainty in this process to be better than ~0.01", or ~40 km at the center of Io's disk, which translates into almost $1.5^0$ deg in latitude/longitude at disk center (increasing with the inverse of the cosine of the emission angle away from disk center). The uncertainty in the positions of hot spots is typically of a similar magnitude (see above mentioned papers), so we assign a total error of $\sim 2^0$ at disk center, increasing towards the limb.

### 3.3 NIRSPEC SO Observations

The high spectral-resolution SO NIRSPEC observations obtained simultaneously with OSIRIS data on 25 December 2015 were discussed in de Kleer et al. (2019a). A similar experiment was conducted on UT 15 April 2019, when Io moved from sunlight into eclipse at UT $13^h:48^m$. During the observations we dithered Io up and down the slit in an ABBA pattern, so that the difference of the two exposures (A-B) provided sky-subtracted spectra of the satellite. By UT 14:45, 14 min before occultation by Jupiter, Io had come too close to the planet to obtain more usable spectra. We obtained a total of six A-B image pairs with echelle setting[7] 1 and another two using echelle setting 2, both in filter 5(H). The wavelength range 1.681 – 1.714 μm was covered in order 45 in setting 1. This range was extended to 1.680 – 1.731 μm by using more orders (44 and 45) and the two slightly different echelle settings (1 and 2). Since we obtained much less data in setting 2, the noise in the final combined spectrum is not constant over wavelength.

The data reduction, including flat-fielding, spatial and spectral mapping, and image rectification, was performed using the REDSPEC pipeline[8], using a combination of arc-lamps (Ar, Kr, Xe, Ne lines) for spectral calibration. The REDSPEC pipeline provided images A-B, which were further processed using custom IDL scripts.

Figure 7a shows the central part of one of the A-B images obtained from the REDSPEC pipeline, with a positive and a negative spectrum of Io. After separating the two spectra in each image, and inverting the negative spectra, they were aligned, median averaged, and corrected for telluric lines by dividing the spectrum by a normalized spectrum of HIP 85755 (also known as c Oph), a 4.8 mag Be star. The final image over the same frequency range as in panel a for echelle setting 1, order 45, is shown in Figure 7b.

We then integrated the intensity at each wavelength in all four median-averaged images (i.e., one image each of orders 44 and 45 of the two echelle settings 1 and 2) over 12 rows (i.e., over 1.55"), centered at the peak emission. This provided four partially overlapping spectra, which were aligned in intensity with order 45, echelle setting 1. The geocentric velocity of Io was ~ -19 km/s during the eclipse (i.e., moving towards the Earth), implying a Doppler shift correction of 0.11 nm. However, since the observed spectra were already perfectly aligned with the model (Section 4), no Doppler shifts were applied (perhaps there was a small imperfection in the wavelength calibration). (Note that the spectral resolution for the OSIRIS spectra is low enough that doppler shifts can be ignored.) The spectra were then combined to give one spectrum ranging in wavelength from 1.680 to 1.730 μm. At wavelengths where spectra overlapped, we used the best (highest SNR) setting/order. We also removed artifacts in the spectrum (such as the vertical bad stripe in Fig. 7)

---

[7] Setting 1: echelle setting of 63.03 and cross disperser equal to 36.88; Setting 2: echelle setting of 62.48 and cross disperser of 36.62.

[8] UCLA infrared lab; http://www2.keck.hawaii.edu/inst/nirspec/redspec

by replacing them with the average of surrounding pixels.

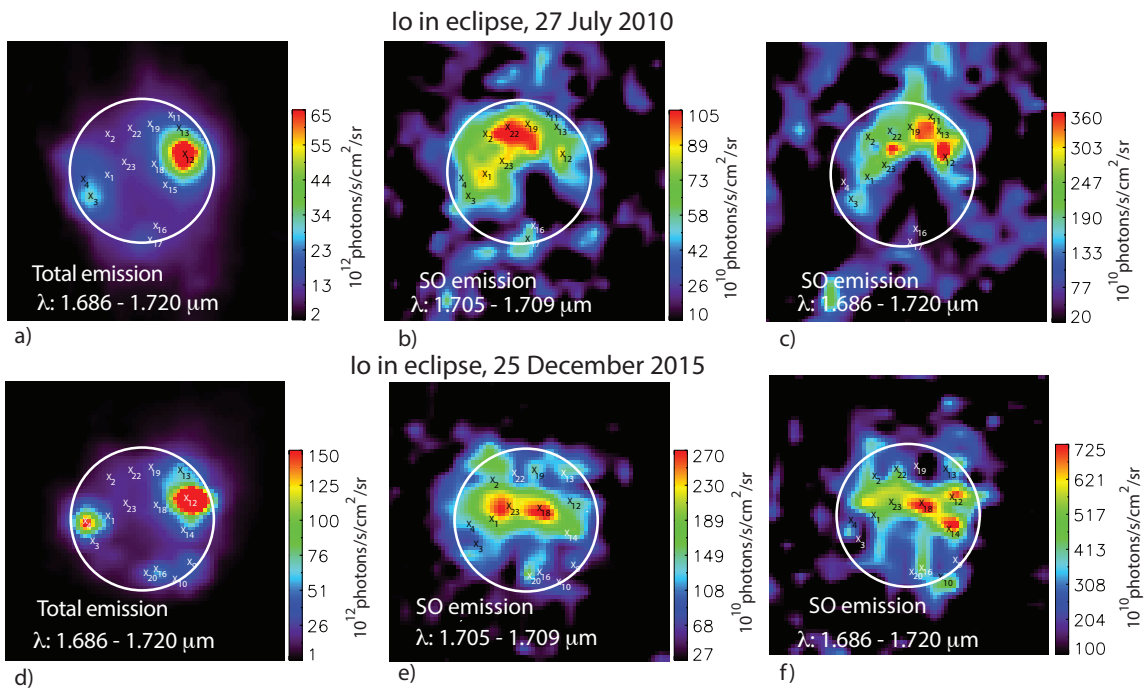

Fig. 6

Fig. 6. a) and d): Images of Io's total emission integrated over the entire SO emission band, including the (background) thermal emission (1.686-1.720 μm) (top row are results from 2010; bottom row from 2015). The bright volcanoes were indicated in Fig. 4. b) and e) Images of Io's emission integrated over the narrow core of the SO emission band (1.705-1.709 μm), after the background had been subtracted from the spectral data-cubes. c) and f) Images of Io's emission integrated over the wings of the SO emission band (1.686-1.720 μm minus the 1.705-1.709 μm range), after the background had been subtracted from the spectral data-cubes. The locations of several volcanic centers are indicated by an x; the sub-scripts refer to the names provided in Table 2 (see Section 4.1 for details).

Calibration was performed in a similar way as for the OSIRIS observations. The NIRC2 observations in the Hcont filter resulted in a total intensity of 4.2 x $10^{-8}$ ergs $s^{-1}$ $cm^{-2}$ μ$m^{-1}$. Converting the July 2010 and December 2015 intensities to the April 2019 epoch, however, shows an intensity of 4.7 x $10^{-8}$ ergs $s^{-1}$ $cm^{-2}$ μ$m^{-1}$. This difference can largely be accounted for by the difference in airmass, which was ~2.4 for the NIRC2 observations in April 2019, and between 1.35 – 1.5 in 2010 and 2015. For a standard H band opacity (τ~0.06) we would need to increase the intensity in 2019 from 4.2 x $10^{-8}$ to 4.5 x $10^{-8}$ ergs $s^{-1}$ $cm^{-2}$ μ$m^{-1}$ to make them consistent with the 2010 and 2015 photometry. Instead of 4.5 x $10^{-8}$ we adopted a total intensity of 4.7 x $10^{-8}$ ergs $s^{-1}$ $cm^{-2}$ μ$m^{-1}$ for Io in the Hcont filter (1.58 μm) (which implies an opacity τ~0.11), or 80% of that at 1.695 μm, i.e., 3.77 x $10^{-8}$ ergs $s^{-1}$ $cm^{-2}$ μ$m^{-1}$.

Although our slit width of 0.72" does not entirely cover Io, by assuming the total intensity of sunlit Io to be equal to 3.77 x $10^{-8}$ ergs $s^{-1}$ $cm^{-2}$ μ$m^{-1}$, we can convert the observed counts/s to a total intensity of the satellite, and use this conversion factor for the Io-in-eclipse spectra. The final spectrum, after smoothing[4] with a Hanning filter and normalizing to a geocentric distance of 4.08 AU (Section 3.2) is shown in Figure 7c. Assuming a blackbody temperature of 675 K as

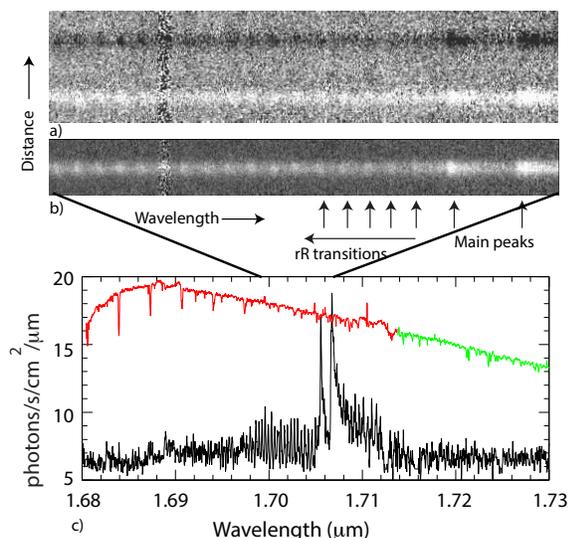

Fig. 7a) Section of a NIRSPEC spectral image from 15 April 2019, order 45 in filter 5(H) with echelle setting 1. This figure shows an A-B image, with a negative spectrum at the top, and positive at the bottom. The wavelength along the x-axis is from 1.699 – 1.707, and distance along the slit is along the y-axis (12" total). b) Final image of order 45, echelle setting 1 after median averaging all 12 spectra. Distance along the y-axis is 6". Note that the two main peaks as well as the rR transitions of the emission band are visible on both panels a and b, though the SNR is considerably improved in panel b.

c) Final NIRSPEC spectrum over the entire wavelength range. The spectrum has been Hanning smoothed like the OSIRIS data, and scaled to a distance of 4.08 AU as all disk-integrated spectra. Superposed is a stellar telluric spectrum (not smoothed), in arbitrary units. The wavelength coverage in red is for order 45, setting 1; green is for order 44, setting 2. Panels a and b show the image of the central section of the spectrum, as indicated.

for the 2010 and 2015 observations, the continuum level of 6.2 photons $s^{-1}$ $cm^{-2}$ $\mu m^{-1}$ corresponds to an effective emitting area of 85-90 $km^2$. In subsequent figures we have subtracted this continuum emission.

### 3.4 Ganymede in Eclipse

We obtained four 30-sec frames of Ganymede-in-eclipse on 25 December 2015, while using Io-in-sunlight for wavefront sensing. After aligning, co-adding (averaging image data-cubes), and averaging the final image data-cube over wavelength, the satellite was clearly visible despite being in Jupiter's shadow, as shown in Figure 8a. The total intensity of Ganymede-in-eclipse is $(7\pm1) \times 10^{-12}$ ergs $s^{-1}$ $cm^{-2}$ $\mu m^{-1}$ (at $\Delta=5.152$ AU), or ~7 mJy. Panels b and c show disk-integrated spectra of Ganymede. Panel b shows the total signal from the satellite, while panel c shows the signal after subtraction of the continuum. The continuum was obtained by averaging the image data-cube over wavelengths outside of the SO band (panel a). Both spectra show that Ganymede does not emit detectable SO emissions.

Due to Ganymede's proximity to Io, Ganymede received a flux from Io (assuming zero phase angle) that is $1.34 \times 10^6$ larger than Io's flux received on Earth. However, due to the large phase angle under which Ganymede sees Io, ~168°, the flux from Io received by Ganymede is ~$10^3$ down from that seen near 0° phase angle (Simonelli and Veverka, 1984), which makes it ~$5 \times 10^{-5}$ ergs $s^{-1}$ $cm^{-2}$ $\mu m^{-1}$. This number will be enhanced by Io's thermal emission. If we assume a total thermal emission equal to that observed from Io-in-eclipse, enhanced by $1.34 \times 10^6$ due to the difference in distances between Io with the Earth and Io with Ganymede, we get a flux of ~$1 \times 10^{-5}$ ergs $s^{-1}$ $cm^{-2}$ $\mu m^{-1}$, which increases Ganymede's total flux density only slightly. Hence the flux density received on Earth from Ganymede due to Io-shine, assuming a perfect 100% reflectivity, will only be $2.3 \times 10^{-15}$ ergs $s^{-1}$ $cm^{-2}$ $\mu m^{-1}$, i.e., roughly a factor of 3000 less than was received. Such large discrepancies between expectations and observations were also seen and discussed by Tsumura et al. (2014). They attributed Ganymede's (and other Galilean satellites') glow during an eclipse to forward scattered sunlight by hazes in Jupiter's upper atmosphere. Our observations support their hypothesis.

### 3.5 $SO_2$-ice Map

On 27 and 28 July 2010 we obtained OSIRIS image data-cubes of the $3\nu_1+\nu_3$ $SO_2$-ice band in the KN2 filter. These data were treated in the same way as in Laver & de Pater (2008; 2009). Because they were centered at different longitudes than in Laver & de Pater, by combining the two sets of maps we are now able to construct a complete $SO_2$-ice map at 2.1258 μm. This map, together with a map of the reflectivity at 2.1 μm and a Voyager visible-light map are shown in Figure 9. Superposed are the locations of all hot spots reported by Cantrall et al. (2018) and de Kleer et al. (2019b).

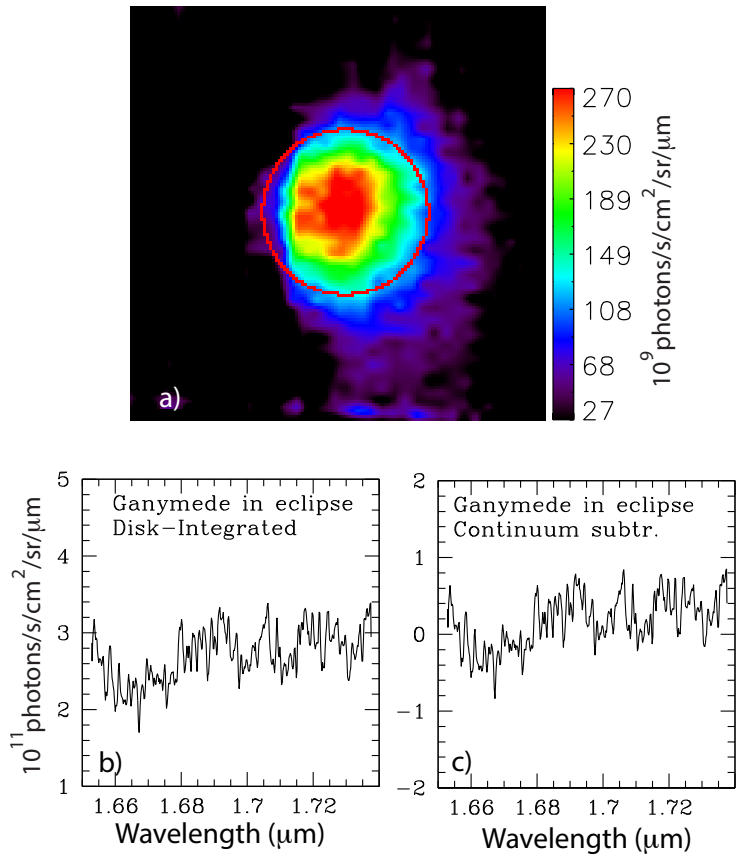

Fig. 8

Fig. 8. a) Image of Ganymede-in-eclipse, averaged over the HN4 band. A circle the size of Ganymede's disk has been superposed. b) Disk-integrated spectrum of Ganymede-in-eclipse. c) Disk-integrated spectrum of Ganymede in panel b after subtraction of the continuum emission.

The 2.1 μm reflectivity shows a good, though not perfect, resemblance to the Voyager map. Dark patera on the Voyager map are typically also dark in the near-infrared, while the bright areas usually show a higher infrared reflectivity. The $SO_2$-ice (equivalent width) map shows a distribution that is concentrated near the equator, which is consistent with the findings by McEwen et al. (1988) and with the *Galileo*/NIMS equivalent width maps at 2.79 and 3.35 μm by Carlson et al. (1997), but differs from Carlson et al.'s equivalent width map at 3.77 μm and with the maps published by Douté et al. (2001). The latter maps were the products of an analysis of *Galileo*/NIMS spectra via a spectral inversion technique, in contrast to equivalent width maps which give essentially a direct representation of the data.

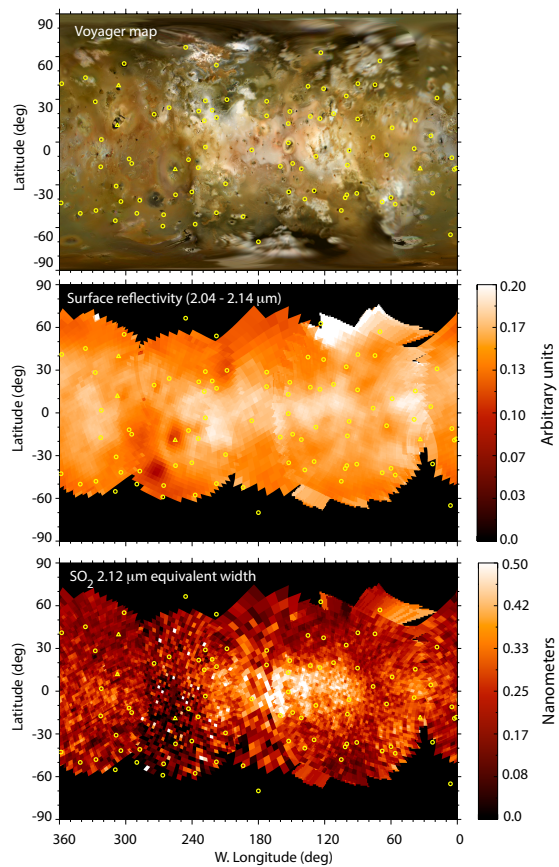

Fig. 9.

Fig. 9. a) Voyager visible light map. b) Surface reflectivity map at 2.04—2.14 μm (in arbitrary units). c) $SO_2$ ice map, shown in the form of the 2.12 μm equivalent width (in nm). Superposed are the locations of all hot spots reported by Cantrall et al. (2018) and de Kleer et al. (2019b). The centers at Loki Patera, Amaterasu Patera, Kanehekili Fluctus, and Pele are indicated by triangles, and all other centers by circles.

Several authors have suggested that the various absorption bands may display differences in sensitivity to the size of frost grains (Schmitt et al., 1994; Carlson et al., 1997; Douté et al., 2001;

Laver and de Pater, 2009), where the weakest (1.98 and 2.12 μm) ice bands are sensitive only to large-grained (> 700 μm) ice deposits, and the stronger absorption bands are quite sensitive to small grain sizes. The strong 3.77 and 4.07 μm absorption bands are even sensitive to thin (few mm) veneers of micron-sized grains, which appear to be abundant at the higher latitudes. Geissler et al. (2001) and Laver & de Pater (2009) explain that the formation of coarse-grained $SO_2$ snowfields near the equator and thin veneers of small-grained frosts at higher latitudes result from a combination of sublimation (at low latitudes), condensation (at latitudes > 27°), and thermal annealing (at low latitudes).

Table 2: Volcanic Identifications

| Number | Name | W. Longitude[a] | Latitude[a] | Past Plume activity[b] |
|---|---|---|---|---|
| 1 | Karei Patera | 13.2 | 0.2 | Yes |
| 2 | Ukko Patera | 18.5 | 33.5 | Yes |
| 3 | Kanehekili Fluctus | 31.9 | -18.5 | Yes |
| 4 | Janus Patera | 37.7 | -3.5 | No |
| 5 | Masubi Fluctus | 54.1 | -44.2 | Yes[c] |
| 6 | Pele | 254.7 | -18.4 | Yes |
| 7 | Svarog | 270 | -51.5 | yes |
| 8 | Daedalus | 267.8 | 20.8 | Yes |
| 9 | Ulgen Patera | 289.8 | -38.2 | No |
| 10 | North Lerna | 290.6 | -56.6 | Yes |
| 11 | Dazhbog | 301.9 | 54.0 | Yes |
| 12 | Loki Patera | 306.5 | 16.0 | Yes[d] |
| 13 | Amaterasu Patera | 303.3 | 39.4 | Yes |
| 14 | Mazda Patera | 310 | -10 | |
| 15 | Ra Patera | 325 | -8 | Yes |
| 16 | PV170 | 329.3 | -46.0 | No |
| 17 | Hiruko Patera | 328.9 | -65.1 | |
| 18 | Acala Fluctus | 334.6 | 9.0 | Yes |
| 19 | Surt | 335.6 | 43.4 | Yes |
| 20 | Creidne Patera | 341 | -50 | Yes |
| 21 | Euboea Fluctus | 355 | -48.8 | Yes |
| 22 | PFU1063 | 357.0 | 40.0 | No |
| 23 | Fjorgyn Fluctus | 358.8 | 10.9 | Yes |

[a] W. Longitude and Latitude (planetocentric) in degrees from Cantrall et al. (2018), de Pater et al. (2014), de Kleer et al. (2019b), or the USGS Io map (https://planetarynames.wr.usgs.gov/).
[b] Plume activity: direct observations, or inferred from surface deposits. Geissler et. al. (2004); Lopes et al. (2007); Spencer et. al. (2008).
[c] Plume detected N and S of Masubi.
[d] Plume was detected at Loki, NE of Loki Patera.

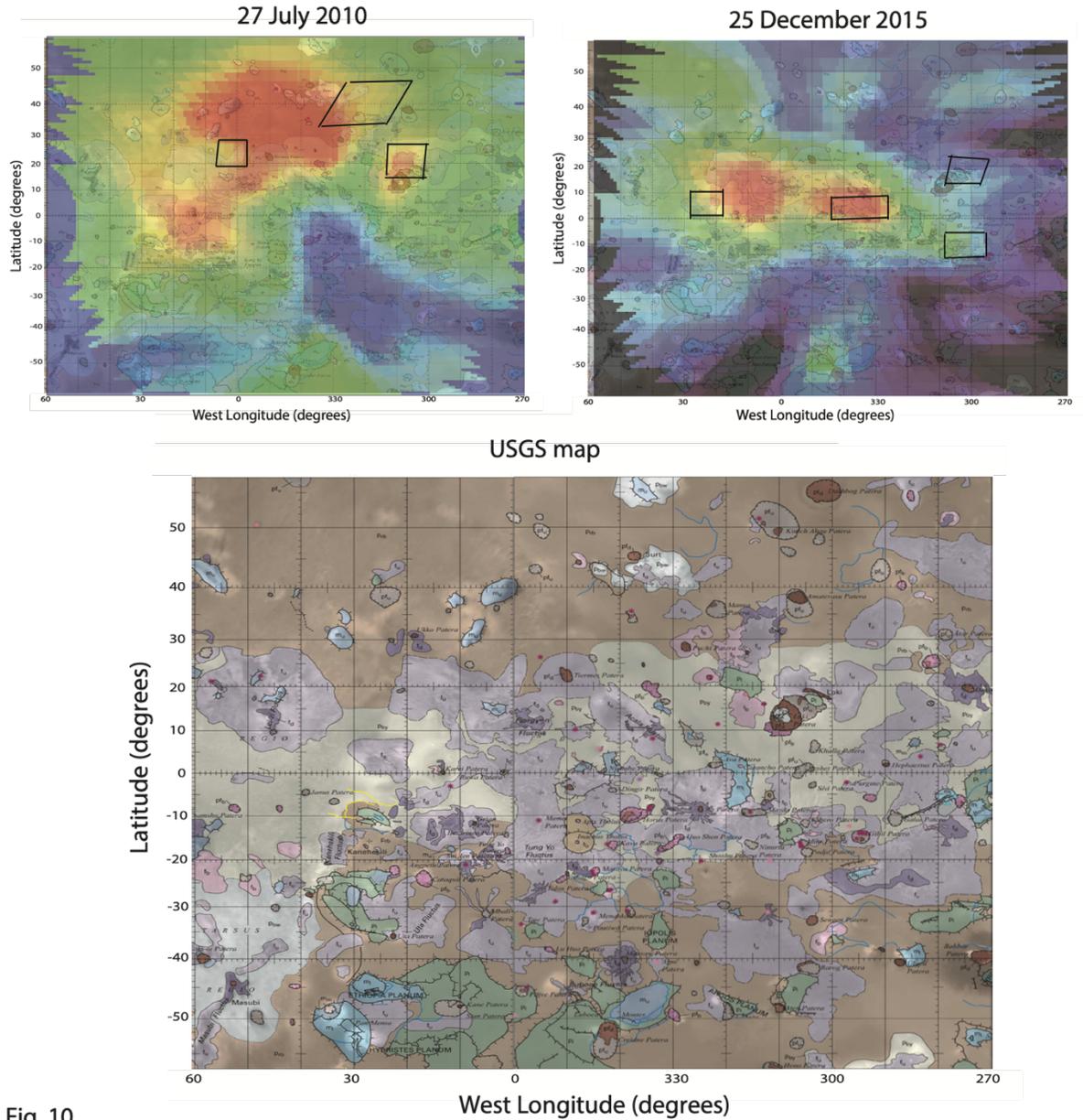

Fig. 10. Reprojected map of the SO emission from Fig. 6b, and 6e superposed on a portion of the USGS geologic map of Io from Williams et al. (2011). The red patches in panels 6c and 6f are indicated by black lines (open boxes) on the figure. The reader is referred to Williams et al. for details on the USGS map.

# 4. Data Analysis & Discussion

## *4.1 Spatial distribution of SO Emission*

Figure 6 shows that the spatial distribution of SO changed between the two epochs[9], while neither epoch shows a strong correlation with the bright volcanic hot spots in panels a and d (and Figure 4). To investigate the correlation between SO emissions and volcanic centers in more detail, we indicated the locations of several volcanic sites on the panels (each indicated by an x); a legend to this Figure is provided in Table 2. This Table lists all the volcanic sites on the observed hemisphere where plumes had been detected in the past, in addition to a few volcanic hot spots seen in our continuum maps or potential candidates for some of the SO emissions. Typical uncertainties in the location of these sites are of order ~2-5$^0$ (Section 3.2).

In several instances the peak SO emission does coincide with a volcanic center, but there certainly is not a clear one-on-one correspondence between patches of SO emissions and known volcanoes. A good correspondence between a volcanic site and SO emission in the core of the emission band (Figs. 6b, e) is seen, e.g., near Loki ($x_{12}$), Karei Patera ($x_1$), Fjorgyn Fluctus ($x_{23}$), and Hiruko Patera ($x_{17}$) in 2010, and Acala Fluctus ($x_{18}$), Surt ($x_{19}$), and Creidne Patera ($x_{20}$) in 2015. Sometimes the wings of the emission band (Figs. 6c, f) are highly suggestive of a volcanic source, such as near Loki ($x_{12}$) in both epochs, and near Mazda Patera ($x_{14}$), N. Lerna ($x_{10}$) and over Acala Fluctus in 2015. Plumes have, or may have, been present in the past at several of these locations (Table 2). Although the very bright SO emission patch in 2010 (panel b) is surrounded by Ukko Patera ($x_2$), PFU1063 ($x_{22}$) and Surt ($x_{19}$), and the western bright patch in 2015 (panel e) is surrounded by Karei Patera ($x_1$), Ukko Patera ($x_2$), and Fjorgyn Fluctus ($x_{23}$), there is no good correspondence between these bright SO patches and one unique volcanic site. In contrast, the eastern patch of SO emissions in 2015 is clearly co-located with Acala Fluctus.

In Figure 10 we superpose a reprojected map of the SO core emission on a USGS geologic map of Io (Williams et al., 2011). The approximate outline of the strong SO wing emissions (red patches in Figs. 6c, f) is indicated by black boxes; these are usually offset from the location of the core of the emission band, except for Acala Fluctus in 2015, and north of Loki Patera in 2010. In 2015 the SO wing emission is strong in the region north-east to the horse-shoe-shaped patera (lake) itself, at Loki, while not much core emission is visible. The Voyager spacecraft detected plumes in this region in 1979 (McEwen et al., 2004). Figure 10 further shows that the large area of strong SO (core) emission in 2010 is located over an area characterized as "red-brown plains", bordered by structures filled with undivided mountain ($m_u$) and flow ($f_u$) materials, white plains material ($P_{bw}$), and undivided ($P_{fu}$) and dark ($P_{fd}$) patera floor materials, indicative of eruptions in the (distant) past. In 2015 the bright SO emissions are also in areas near dark ($f_d$) and undivided ($f_u$) flow materials, as well as areas overlain by white bright plains material, i.e., regions dominated by $SO_2$ frost.

A comparison with the $SO_2$ ice map in Figure 9 does not reveal an obvious connection of SO emissions to $SO_2$ frost, although in 2015 the large patches of SO emissions are within ~15-20$^0$ of the equator, where we expect relatively thick large-grain ice deposits based upon our 2.12 μm observations; this is also a region of bright plain deposits as shown in Figure 10. The large area of SO emissions in 2010, centered near 350° W. longitude, 35° N. latitude coincides with the small-grained $SO_2$-ice deposit revealed by Douté et al. (2001). Unfortunately, the western patch of SO emissions in 2015 is over a region not mapped by Douté et al. (2001). We conclude from these comparisons that many of the SO emissions not associated with known volcanic sources appear to be located over regions of $SO_2$ frost.

---

[9] The main difference between our 2010 and 2015 observations is the geometry (Table 1); we do not think this could translate into such a difference in the locations of the SO emissions.

Although it is possible that SO upon escaping a volcanic vent is redistributed spatially by winds, the connection of some SO locations to known volcanic sites, and in particular several with known past plume activity, argues against winds redistributing the volcanic gases. The figures, we think, are highly suggestive of more than a single compact source of SO, such as the presence of a (perhaps large) number of "stealth" plumes, an idea originally suggested by Johnson et al. (1995) to explain the patchiness in the $SO_2$ atmosphere as inferred from UV and microwave observations. The authors suggest that, in contrast to the "low-to-moderate entropy" Prometheus plume (Kieffer et al., 2000), stealth plumes are "high-entropy" eruptions from a reservoir of superheated $SO_2$ vapor in contact with silicate melts about 1.5 km below the surface at pressures of ~40 bar and temperatures of ~1400 K. Such plumes would consist of essentially pure gas, i.e., without dust or condensates, so that they cannot be detected in reflected sunlight. Such plumes can, and have been, detected during eclipse observations, such as the plumes and diffuse glows that were imaged by the *Galileo* spacecraft over Acala Fluctus (McEwen et al., 1998). Faint glows and numerous tiny point source emissions were also detected in eclipse images by the *New Horizons* mission in the general area where we see the bright SO patches in 2015 (Spencer et al., 2008). No specific gas emissions could be specified, however, in those broad-band (400-900 nm) Lorri images. The authors interpreted the emissions as being caused by nonthermal, likely plasma-related, near-surface processes. In contrast to the *Galileo* and *New Horizons* data, however, we do not see emissions on the limb of Io, as might be expected if the emission process is related to magnetospheric plasma processes.

In the original paper on the detection of the forbidden SO 1.707 μm rovibronic transition ($a^1\Delta \rightarrow X^3\Sigma^-$) on Io (de Pater et al., 2002), the authors show that the SO electronic states are equilibrated at a quenching temperature of ~1500 K, which, as expected, is well above the rotational temperature of the gas. Since they could not explain the emissions in any other way (e.g., through electron impact, Joule heating, ionospheric recombination), they stated: "The only plausible explanation for the observed SO emissions is direct ejection of excited SO from the volcanic vent." They suggested Loki as its source, which was particularly bright at infrared wavelengths at the time. In the present paper we suggest that excited SO is ejected directly from a large number of stealth volcanoes, where temperatures at depth are of order 1400 K. These emissions cover the large patches of SO emissions we see in Figures 6 and 10, which includes Acala Fluctus. In addition, as shown by a direct correlation with some volcanoes (e.g., Loki, Karei Patera, Fjorgyn Fluctus, Surt), excited SO must be directly ejected from most active volcanoes.

### 4.2 Spectral Shape of OSIRIS SO Emission

In order to learn more about the characteristics of the SO emission, such as the temperature(s) of the emitting gas and potential non-LTE effects in addition to the spatial variations discussed above, we constructed spectra integrated over several small and a few larger areas, as indicated graphically on Figure 11. The spectra integrated over the large rectangles are shown in Figure 12; those over the small ones in Figure 13.

We modeled the SO emission assuming local thermodynamic equilibrium (LTE) as in de Pater et al. (2002) and Laver et al. (2007). As shown by the latter authors, the width of the core of the emission band increases with the rotational temperature. Best fits to the central part or core of the spectra (1.705 – 1.709 μm) are shown by red lines in Figures 3 and 12; these fits were obtained from the minimum values in chi-square fits[10], as shown graphically for the disk-integrated spectra in the bottom row of Figure 3. Such graphs can also be used to derive the uncertainty in these fits: by doubling the chi-square minimum values, we find a rotational temperature for the disk-integrated spectra of $350^{+400}_{-150}$ K in 2010, and $500^{+150}_{-125}$ K in 2015. The middle row in Figure 3 also shows profiles for rotational temperatures of gas 200 K warmer (blue lines) and colder (cyan lines) than the best-fit curves. In Figure 12 we only indicate the best fit temperatures. Lower limits to these values are typically 100-150 K below the best-fit values, and upper limits are usually 200-400 K above the indicated values; the errorbars are quite sensitive, though, to the precise region that is modeled. Although, as shown, the center part of the emission bands can be matched well with these LTE models, it is clear that none of these models match the extended structure ("shoulders" or wings) of the emission bands, in particular near 1.69 μm; this mismatch is a clear indication of multiple temperatures and/or non-LTE effects.

---

[10] Calculated as: $[\Sigma_i (obs_i - model_i)^2]/\sigma^2$, summed over all points *i* in the narrow core of the emission line; σ is the standard deviation in the spectrum.

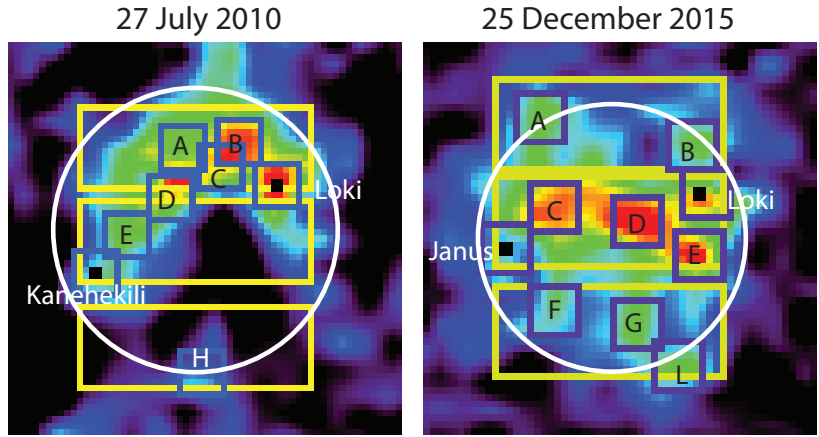

Fig. 11

Fig. 11. Images similar to those in Fig. 6, but integrated over the full SO emission band (1.686-1.720 µm), at an arbitrary color scale. These images were created by interpolating the original data on a grid that was 4 times larger than the original image (the images in Figures 4, 6 were processed in the same way). Superposed are small annotated squares (8x8 pixels), and larger rectangles (38x14 pixels in 2010, and 34x14 pixels in 2015). The bright volcanoes seen in the continuum map (Loki Patera and Kanehekili Fluctus in 2010; Loki Patera and Janus Patera in 2015) are indicated by solid black squares within the small square. Spectra integrated over these rectangles and squares are shown in Figs. 12 - 14.

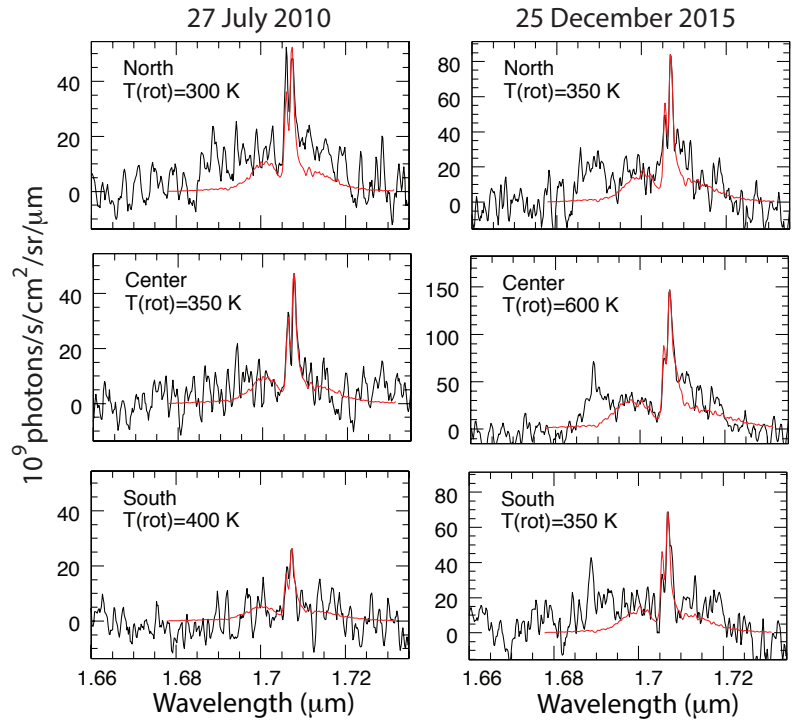

Fig. 12

Fig. 12. Spectra integrated over the yellow rectangles in Fig. 11, indicated by North, Center, and South. Superposed (in red) are LTE models of the SO emission that best fit the center portion or core of the emission band; the rotational temperatures of these lines are indicated on each panel.

Spectra of smaller regions (small boxes in Fig. 11) are shown in Figure 13. Despite the relatively low SNR in the spectra, one can notice several interesting phenomena, in particular when also comparing the spectra with the spatial distribution of emissions in Figures 6 and 10. In 2010, the core of the emission band is most clearly detected in regions A (SNR~7.1, as determined from Fig. 6), C (SNR~6.8), D (SNR~5.8) and E (SNR~5.6), and can be distinguished at Loki (SNR~5.5), Kanehekili (SNR~4) and perhaps in region H (SNR~3.5). In some regions the wings of the emission band are quite bright; we note in particularly Loki (SNR~7.5) and region B (SNR~5.6), where the wings are clearly present, but the core of the emission band can hardly be distinguished above the wings. In 2015 the core of the emission band is clearly visible near Janus Patera (SNR~6), and regions A (SNR~8.2), C (SNR~13.5), and D (SNR~13.5), while the wings are visible at several locations where the core of the emission band is not or only faintly visible above the wings, such as at Loki (SNR~7.3), E (SNR~7.8), F, G, and L (each with SNR~4-5). The emission bump at 1.69 μm is particularly strong in regions C and D in 2015, but is visible at several other locations as well (e.g., B, C in 2010; Loki, Janus,

A and E in 2015). The narrow emission band cores in most of these spectra are suggestive of temperatures of order a few 100 K, while the rotational temperature for regions C and D in 2015 are of order 600 K, just as for the Center region in Figure 12. Clearly, there is a lot of heterogeneity in both the shape and the spatial distribution of the SO emissions.

De Kleer et al. (2019a) analyzed the 2015 high spectral resolution data obtained with NIRSPEC on Keck II at the exact same time as we observed with OSIRIS. In order to fit their data, they adopted a gas population where the high and low rotational levels are populated according to Boltzmann distributions at a high and a low temperature. They obtained a best fit to the spectra using $c_1 F(T_1) + c_2 F(T_2)$, with F the model intensity at temperature $T_1$=186 K and $T_2$=1500 K, and the ratio $c_2/c_1 = 5/6$[11]. In Figure 14 we superpose essentially the same model on our OSIRIS 2015 spectra, normalized to the peak intensity of each spectrum. Panel a) shows the disk-integrated SO profile as determined from the final spectral image data-cube. Panel b) shows the center part of Io (yellow box in Fig. 11) and panels c) and d) show spectra integrated over the small boxes C and D in Figure 11. As shown, the model fits both the core of the emission band and the wings quite well, except for the 1.69 μm bump. This bump, though, fell outside of the 2015 NIRSPEC wavelength coverage, which was between 1.694 and 1.717 μm. However, it seems quite impossible to adapt the model to fit this emission bump.

To gain a better understanding of the emission line complex, we show transitions grouped by branch in Figure 15 for a temperature of 300 and 1500 K. The transitions covering the extended emission branch increase at the higher temperatures. Indeed, the shoulders of the SO emission can be more or less matched by increasing the temperature, as shown by the cyan line in Fig. 14b. However, an increased temperature broadens the main component of the line profile, which has not been observed. The combination of two temperature profiles, a low and a high T, solves this problem to some extent, as shown by the three curves in Fig. 14b. However, all these combinations fail to show enhanced emission at 1.69 μm. This can only be brought about by adding a gas component at a high temperature, and including only high rotational states, such as $J > 50$ at 1800 K, shown in Figure 14a. This is a pure thought-experiment, though, as it seems physically implausible to only excite the high J states, or collisionally de-excite only the low J states.

---

[11] We note that in Figure 16 of de Kleer et al. (2019a), the inverse of $c_1$ and $c_2$ is plotted, rather than $c_1$ and $c_2$.

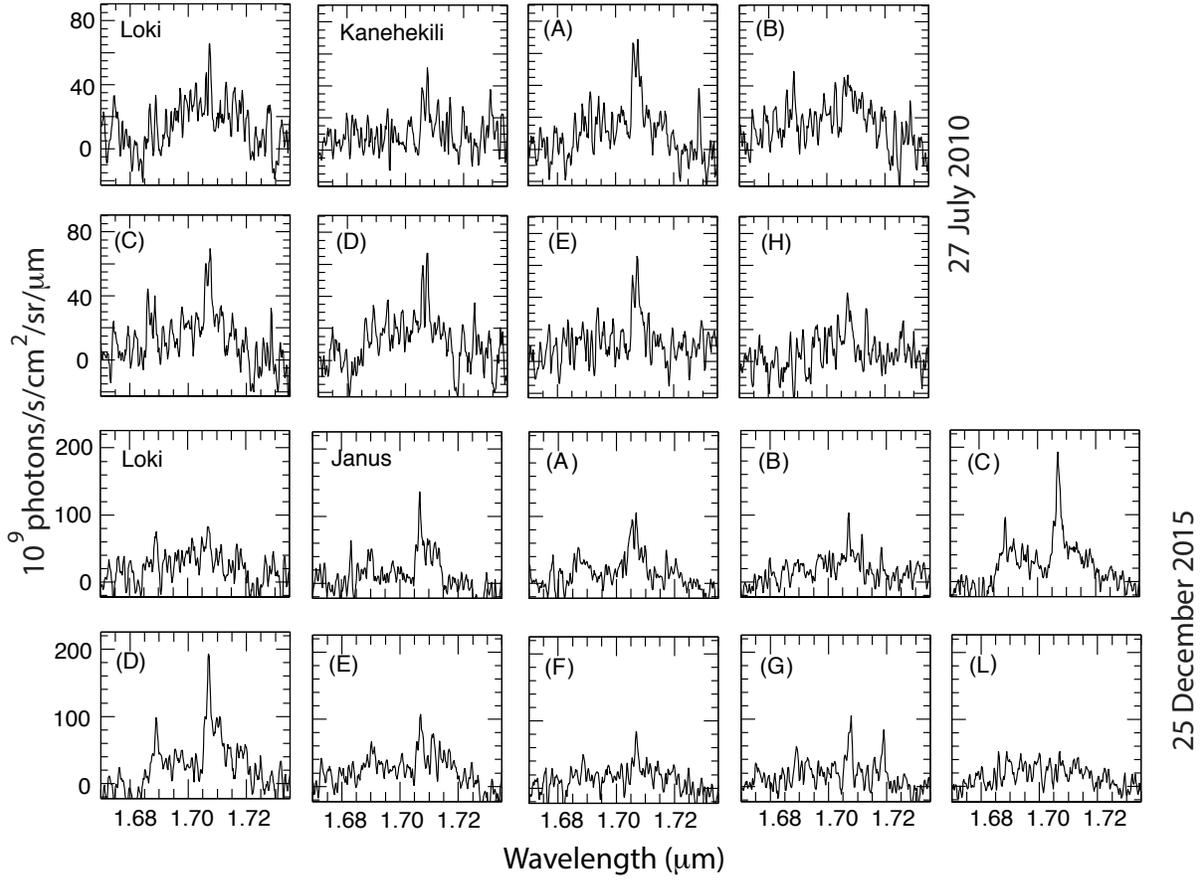

Fig. 13

Fig. 13. Spectra integrated over the small squares in Fig. 11. The 2 top rows show spectra from 2010; the 2 bottom rows from 2015.

### *4.3 Spectral Shape of April 2019 NIRSPEC SO Emission*

As shown in OSIRIS spectra (e.g., Fig. 14), we often see an emission bump near 1.69 μm. This emission feature has been seen in most of our older NIRSPEC observations at a similar (i.e., medium) spectral resolution as our OSIRIS data (de Pater et al., 2002; Laver et al., 2007). The feature is at times very prominent; it clearly is variable both in time (based upon this paper and previous measurements) and with location (this paper) on Io's disk. This wavelength range was, unfortunately, not covered by our 2015 high spectral resolution NIRSPEC data. To remedy this shortcoming, we observed Io-in-eclipse again with NIRSPEC at high spectral resolution on 15 April 2019, using a spectral setting that did cover the 1.690 emission bump. These data were shown in Figure 7c. In Figure 16a we show the central part of the line with superposed a best fit model using the procedures from de Kleer et al. (2019a), normalized to the peak intensity in the data (in red). The best fit temperatures to the 2019 data are $T_1$=100 K and $T_2$=1120 K, and the ratio $c_2/c_1$ = 1.10. As shown in panel b, some emission was observed near 1.69 μm, but this is absent in the model.

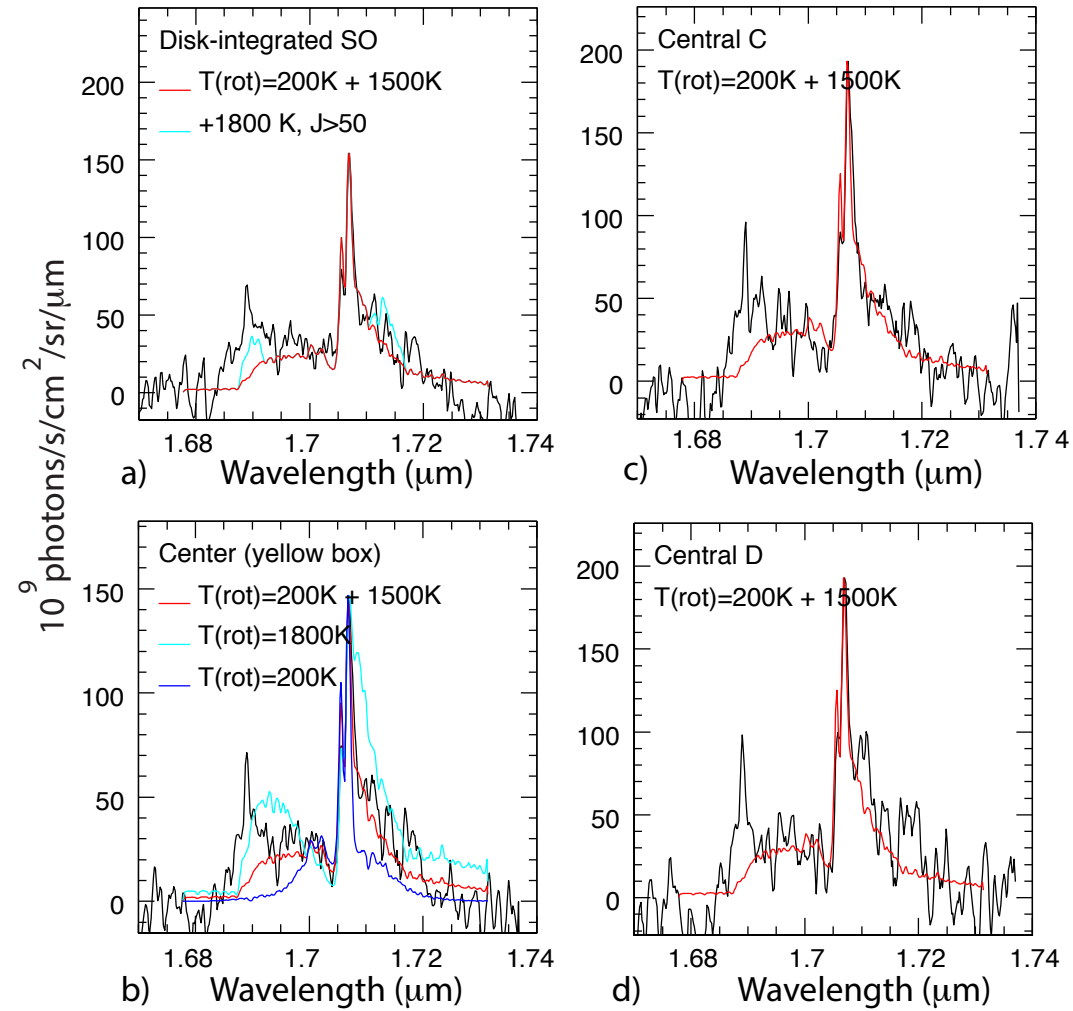

Fig. 14

Fig. 14. High SNR OSIRIS spectra from 25 Dec. 2015, with superposed (in red) a model after de Kleer et al's (2019a) best fit model to NIRSPEC's 2015 high-spectral resolution data (normalized to the peak intensity of each spectrum). This model consists of a gas with 2 temperatures: $T_1$=200 K and $T_2$= 1500 K, in almost equal proportions ($c_2/c_1$=6/5). Panel a shows the total flux density; panel b shows the center panel from Fig. 12 (right column), and panels c and d the spectra from Fig. 13. In panel b we also show a profile for single rotational temperatures of 200 and 1800 K, and in panel a we show the contribution of only high J-states at 1800 K to the 2-temperature profile (cyan).

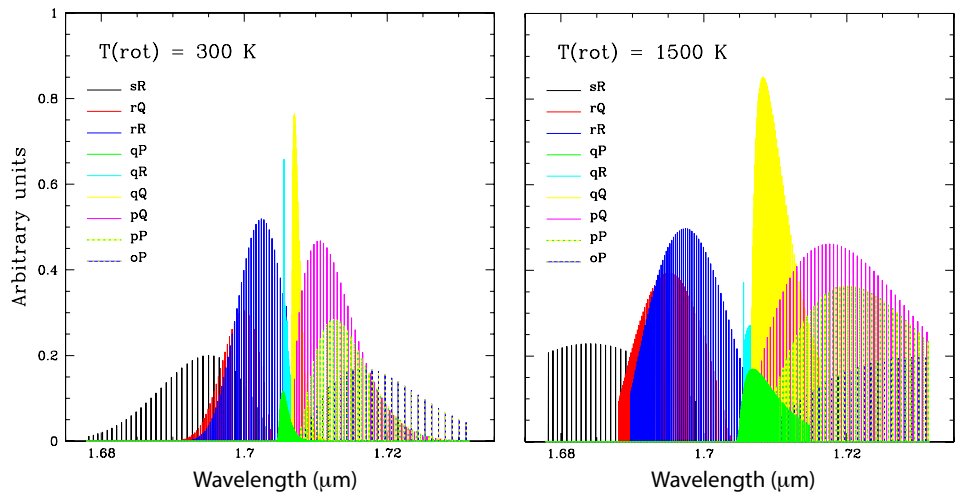

Fig. 15

Fig. 15. The various transition branches of SO at 300 K (left) and 1500 K (right). Note the relative changes that take place when the temperature is increased.

Figure 16c shows the entire 2019 NIRSPEC spectrum smoothed down to the approximate resolution of the OSIRIS data (black line). Superposed is the above model, smoothed down to the same spectral resolution (in red). The 2015 disk-integrated OSIRIS spectrum from Fig. 14a, normalized to the NIRSPEC spectrum, is also superposed. The 2015 OSIRIS, 2019 NIRSPEC and the de Kleer model agree quite well at wavelengths 1.698—1.712 μm. At longer wavelengths the NIRSPEC spectrum was taken with Echelle setting 2 (Fig. 7c), where the SNR was much lower. The drop in NIRSPEC intensity at 1.712 μm is caused by bad data (see Fig. 7c). Shortwards of 1.698 μm the difference is caused by the 1.69 μm bump. Note that a small bump at this wavelength is visible in the 2019 NIRSPEC data. As mentioned above, this bump varies both in time and location on the disk, and hence the fact that we do see differences in strength should not be unexpected. The model, though, does not match either bump.

Recently, Bernath and Bittner (2020) presented a new line list for the $a^1\Delta \rightarrow X^3\Sigma^-$ transition. They provide line lists for both the 0-0 and 1-1 bands. In Figure 17 we compare our data with these new models. The results for matching only the 0-0 band (as in de Kleer et al's work) is very similar to the model in Figure 16a,b. The model in Figure 17a,b shows the best fit for the combined 0-0 and 1-1 bands. The temperatures were quite similar as before: $T_1$=80 K and $T_2$=1120 K. However, the ratio $c_2/c_1 \sim \frac{1}{4}$, i.e., roughly a factor 4 lower than before. The separate 0-0 and 1-1 band model components are also shown. We note that the emission bumps in the model near 1.717 μm (qR13 and qQ12) are not visible in the data, while the 1.69 μm emissions in the data (rQ1, rR12) are not obvious in the model.

Figure 17c shows smoothed profiles, as in Figure 16c. The new model does not match the observed emission bumps near 1.69 μm, while the 1.717 μm emissions in the 1-1 band are not obvious in the data (although it might be weakly present in the OSIRIS spectrum). The 1.69 μm bump, or at least elevated temperatures at 1.69—1.70 μm, are visible in high-temperature models. The magenta curve is a spectrum for a model with a temperature of 1600 K, normalized to the observed peak flux density. In addition to the elevated temperatures at 1.69—1.70 µm, this curve shows strong 1-1 band qR13 and qQ12 emission bumps near 1.717 µm, which are not as apparent in any of the data. The model is also much too broad in the core of the line (see also Fig. 14b). Since the strong qQ12 component of the emission in the 1-1 band had not been seen in past observations, and the 1.69 µm emission bump (rR12) is much stronger in the data compared to their model, Bernath and Bittner (2020) concluded that there is a lack of thermodynamic equilibrium in the Io emissions. Their statement is confirmed by the data presented in this paper. Hence, in addition to our conclusion that much of the observed SO emission is caused by direct ejection of excited SO molecules from a large number of stealth plume volcanoes, the data also indicate the presence of non-LTE processes, perhaps caused by the complicated interaction of volcanic plumes with the atmosphere and surface frost (see, e.g., Zhang et al., 2003).

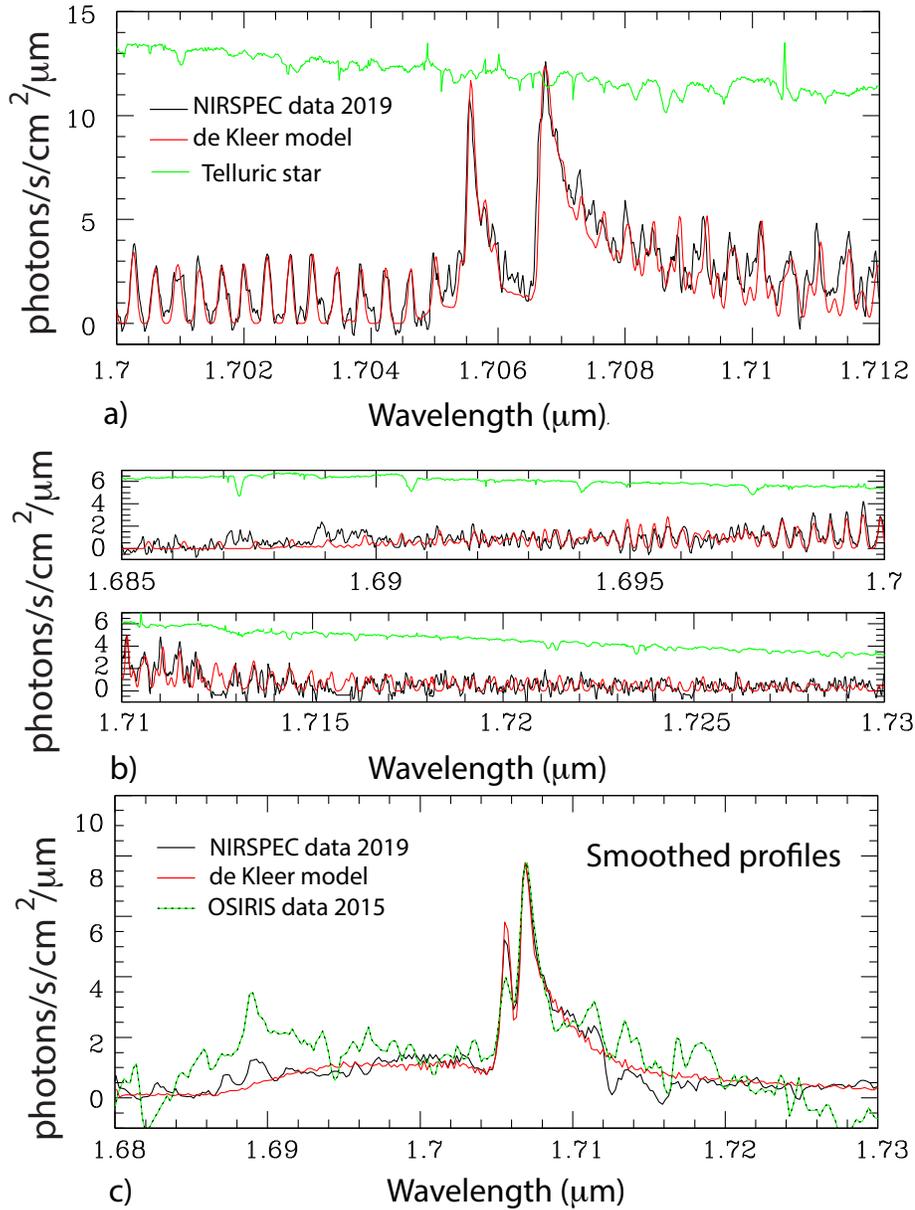

Fig. 16

Fig. 16. a) 2019 NIRSPEC spectrum at 1.700-1.712 mm from Figure 7c, with superposed (in red) a 2-temperature model that best fits the data using the de Kleer et al. (2019a) procedures, normalized to the peak intensity of the data. $T_1$= 100 K, $T_2$ = 1120 K, and $c_2/c_1$~1.10. b) The spectrum at 1.685-1.700 m and 1.710-1.730 m, with superposed the model from panel a. c) 2019 NIRSPEC spectrum smoothed down to the approximate resolution of the 2015 OSIRIS data, with superposed the model from panel a, also smoothed down. The green/black line shows the 2015 OSIRIS disk-integrated data from Fig. 14a, normalized to the peak intensity of the 2019 NIRSPEC data. At the top of both panels a and b stellar telluric spectra from Fig. 7c are shown (in green).

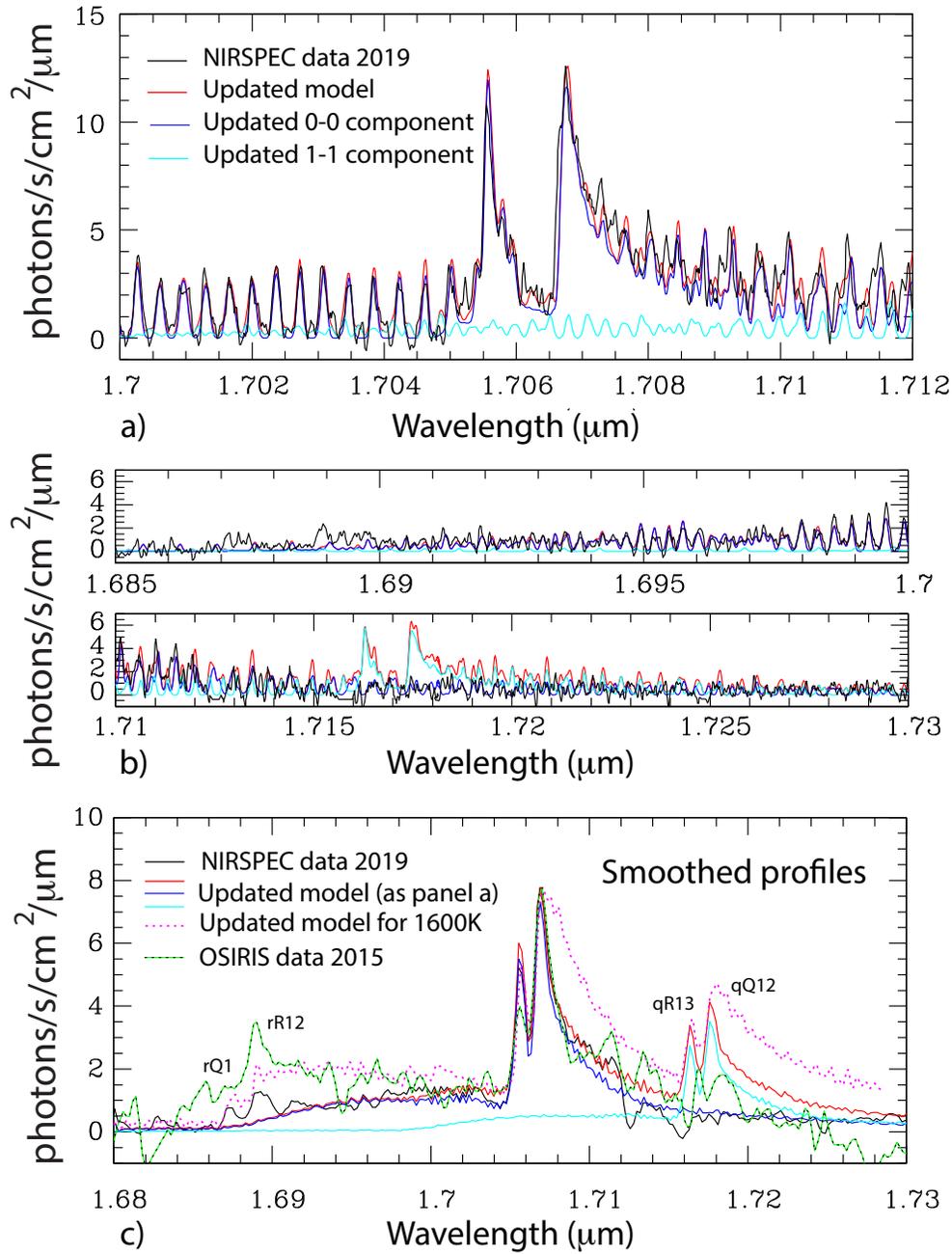

Fig. 17

Fig. 17. Same 2019 NIRSPEC spectrum and panels as in Figure 17, with superposed models based upon the updated line list from Bernath & Bittner (2020). a) SO emission band at 1.70-1.712 μm with superposed (red) a best-fit 2-temperature model, consisting of both the 0-0 and 1-1 band models. $T_1$= 80 K, $T_2$ = 1120 K, and $c_2/c_1 \sim 1/4$. The blue line is for the 0-0 band; the cyan line for the 1-1 band. b) The same spectra as in panel a, at 1.685-1.700 and 1.710-1.730 μm. c) 2019 NIRSPEC spectrum smoothed down to the approximate resolution of the 2015 OSIRIS data, with superposed the smoothed model, as in panel a. The green/black line shows the OSIRIS disk-integrated data, normalized to the peak intensity of the NIRSPEC data. The dotted magenta line shows a single temperature model for T=1600 K.

## 5. Conclusions

We observed the forbidden SO $a^1\Delta \rightarrow X^3\Sigma^-$ rovibronic transition at 1.707 µm with the field-integral spectrometer on the Keck telescope on 27 July 2010 and 25 December 2015 while the satellite was in eclipse; the spatial resolution was ~0.12" and spectral resolution R ~2,500 over a range 1.652-1.737 µm. From simultaneously obtained NIRSPEC spectra over 1.694 – 1.717 µm at a high spectral resolution (R~15,000) on 25 December, de Kleer et al. (2019a) obtained a best fit to the total emission by using two temperatures: $T_1$=186 K and $T_2$=1500 K in almost equal proportions. A similar model with $T_1$=100 K and $T_2$=1120 K matches new high spectral resolution NIRSPEC data obtained on 15 April 2019, covering a wavelength range 1.680-1.731 µm. However, none of these models could not match the emission bump near 1.69 µm, observed in the OSIRIS and new NIRSPEC data, as well as most previous medium resolution NIRSPEC spectra. While writing this paper, and after de Kleer et al.'s (2019a) publication, Bernath and Bittner (2020) published a new line list for the $a^1\Delta \rightarrow X^3\Sigma^-$ transition. This new (updated) model can be matched to the data by decreasing the high-temperature contribution in the model by a factor of ~4 ($c_2/c_1$=1/4), with temperatures $T_1$=80 K and $T_2$=1120 K. This model does not provide a better match to the 1.69 µm emission bump; temperatures of order 1600 K are needed to match this emission bump, but such models do not fit the core of the emission line. The 1-1 band in the new model shows emission bumps near 1.717 µm, which are not obvious in the data. Both the presence of the 1.69 µm and absence of the 1.717 µm emissions in the data led Bernath and Bittner (2020) to conclude that "SO is not in thermodynamic equilibrium", a conclusion we support in this paper.

The main scientific results of our paper can be summarized as follows:

- There is considerable variability in the shape of the SO emission spectrum both across the disk and over time. In addition, the detailed spatial distribution differs between the core of the emission band (1.705-1.709 µm) and the wings.

- The center of the line (1.705-1.709 µm) is indicative of rotational temperatures varying from a few 100 to ~600 K, depending on location. The wings of the emission band are indicative of high (>1500 K) temperatures and non-LTE effects.

- In some cases the SO emissions in the core and/or the wings of the emission band can be identified with volcanoes. Most large SO emission patches, however, do not coincide with known volcanoes or volcanic constructs, with the exception of the eastern complex in 2015, which overlaps with Acala Fluctus. Evidence of past volcanic activity is, however, usually seen around these areas. (SO emissions are seen, e.g., over Loki (N/NE of Loki Patera), Karei Patera, Fjorgyn Fluctus, Surt, Creidne Patera, Mazda Patera and the North Lerna region).

- The large areas of SO emissions in 2015 are located close to the equator where our new $SO_2$-ice maps indicate the possible presence of $SO_2$ ice deposits. In 2010 the emissions are over an area where Douté et al. (2001) detected a thin veneer of micron-sized $SO_2$ ice grains.

- We suggest the large patches of SO emissions to result from a large number of stealth plumes (Johnson et al., 1995), i.e., "high-entropy" eruptions from a reservoir of superheated $SO_2$ vapor in contact with silicate melts at about 1.5 km depth. The emissions are caused by the direct ejection of excited SO from these volcanic vents. The emissions are thus suggestive of widespread stealth volcanism.

- The spectra (in particular the wings of the emission band, the 1.69 μm bump, and absence of the 1.717 μm emissions) and their spatial distribution show signatures of non-LTE processes, confirming the following statement in de Pater et al. (2002): "it is difficult to compete with the fast radiative decay rate of 2.2 s$^{-1}$ (Klotz et al. 1984) to enforce an equilibrium population of electronic states, so LTE might not prevail even near the surface and a volcanic vent."

- On 15 December 2015 we observed Ganymede in eclipse before Io went into eclipse. The observed glow from Ganymede is most likely caused by sunlight scattered in the forward direction by hazes in Jupiter's upper atmosphere, as originally proposed by Tsumura et al. (2014). No SO emissions were detected.

*Acknowledgements*

We thank Darrell Strobel for fruitful discussions regarding the interpretation of our observations. We appreciate Peter Bernath's communication pointing out his recent paper on new line lists of SO. We thank Emmanuel Lellouch and an anonymous referee for careful and detailed reviews of our manuscript, which helped improve the paper substantially. We further thank Edward Molter for reducing the NIRC2 data from 15 April 2019. Our research was supported by the National Science Foundation, NSF grant AST-1313485 to UC Berkeley. The data presented in this paper were obtained at the W.M. Keck Observatories. The Keck Telescopes are operated as a scientific partnership among the California Institute of Technology, the University of California and the National Aeronautics and Space Administration. The Observatory was made possible by the generous financial support of the W.M. Keck Foundation. The authors recognize and acknowledge the very significant cultural role and reverence that the summit of Maunakea has always had within the indigenous Hawaiian community. We are most fortunate to have the opportunity to conduct observations of Ionian volcanoes from this Hawaiian volcano.